\documentclass[journal]{IEEEtran}
\usepackage{caption,comment,threeparttable,multirow}
\usepackage{graphicx}
\usepackage{soul}
\usepackage{cite}
\usepackage{bm}
\usepackage{amssymb}
\usepackage{amsmath}
\usepackage{array,color}
\usepackage{xcolor}
\usepackage[caption=false,font=footnotesize]{subfig}

\definecolor{celadon}{rgb}{0.67, 0.88, 0.69}
\makeatletter
\def\bstctlcite{\@ifnextchar[{\@bstctlcite}{\@bstctlcite[@auxout]}}
\def\@bstctlcite[#1]#2{\@bsphack
\@for\@citeb:=#2\do{%
\edef\@citeb{\expandafter\@firstofone\@citeb}%
\if@filesw\immediate\write\csname #1\endcsname{\string\citation{\@citeb}}\fi}%
\@esphack}
\makeatother
\begin{document}
\bstctlcite{IEEEexample:BSTcontrol}

\title{Encryption-then-Compression Systems using Grayscale-based Image Encryption for JPEG Images}

\author{Tatsuya~Chuman,~\IEEEmembership{}
        Warit~Sirichotedumrong,~\IEEEmembership{Student Member,~IEEE,}
        and~Hitoshi~Kiya,~\IEEEmembership{Fellow,~IEEE}
\thanks{T. Chuman, S. Warit and H. Kiya are with the Department of Information and Communication Systems, Tokyo Metropolitan University,
Tokyo, 191-0065, Japan.}
\thanks{Manuscript received April 6, 2018; revised August xx, xxxx.}}

\markboth{Journal of \LaTeX\ Class Files,~Vol.~xx, No.~x, August~xxxx}%
{Shell \MakeLowercase{\textit{et al.}}: Bare Demo of IEEEtran.cls for IEEE Journals}
%



\maketitle

\begin{abstract}
A block scrambling-based encryption scheme is presented to enhance the security
of Encryption-then-Compression (EtC) systems with JPEG
compression, which allow us  to securely transmit images through an untrusted
channel provider, such as social network service providers.
The proposed scheme enables the use of a smaller block size and a larger number of blocks than the conventional scheme. Images encrypted using the proposed scheme include less color information due to the use of grayscale images even when the original image has three color channels.
These features enhance security against various attacks such as jigsaw puzzle solver and brute-force attacks.
In an experiment, the security  against jigsaw puzzle solver attacks is
evaluated.
Encrypted images were uploaded to and then downloaded from Facebook and
Twitter, and the results demonstrated that the proposed scheme is effective
for EtC systems.
\end{abstract}


%
\IEEEpeerreviewmaketitle

\section{Introduction}
\label{sec:introduction}
%
%
%
%
\IEEEPARstart{T}{he}
use of images and video sequences has greatly increased because of the rapid growth of the Internet and widespread use of multimedia systems. While many studies on secure, efficient, and flexible communications have been reported\cite{huang2014survey,lagendijk2013encrypted,zhou2014designing,2013_Ra}, full encryption with provable security (like RSA and AES) is the most secure option for securing multimedia data. However, there is a trade-off between security and other requirements such as low processing demand, bitstream compliance, and signal processing in the encrypted domain. Several perceptual encryption schemes have been developed to achieve this trade-off\cite{Zeng_2003,Ito_2008,Kiya_2008,Ito_2009,Tang_2014,Chengqing}.
\par
Image encryption prior to image compression is required in certain practical scenarios such as secure image transmission through an untrusted channel provider. Encryption-then-Compression (EtC) systems\cite{zhou2014designing,Erkin_2007,nimbokar2014survey} are used in such scenarios. In this paper, we focus on EtC systems although the traditional way of securely transmitting images is to use a compression-then-encryption (CtE) system.
Most studies on EtC systems assumed the use of a proprietary
compression scheme incompatible with international compression
standards such as
JPEG\cite{Johnson_2004,Liu_2010,Zhang_2010,Hu_2014,zhou2014designing,2018_Liu},
so they can not be applied to social media. For example, an arithmetic
coding-based approach and a singular value decomposition transformation are used
to efficiently compress encrypted images in \cite{zhou2014designing,2018_Liu},
respectively.
Because of such a situation, block scrambling-based image encryption schemes,
which are compatible with international standards, have been proposed for EtC
systems\cite{watanabe2015encryption,kurihara2015encryption,KURIHARA2015,KuriharaBMSB,Kuri_2017}.
However, the conventional block scrambling-based encryption schemes have a
limitation on block size to prevent JPEG distortion.
\par
In this paper, we present a novel block scrambling-based image encryption
scheme for EtC systems that enhances security compared with the
conventional scheme.
Compared with the conventional schemes, for which robustness against several
attacks such as jigsaw puzzle and brute-force attacks has been
discussed\cite{CHUMAN2017ICASSP,CHUMAN2017ICME}, the proposed one enables the
use of a smaller block size and a larger number of blocks, which enhances both
invisibility and security against several attacks. Furthermore, images encrypted
by using the proposed scheme include less color information due to the use
of grayscale images, which makes the EtC system more robust. Although EtC
systems can be applied to social media by using JPEG
images\cite{CHUMAN2017APSIPA}, there is a limitation on block size to prevent
JPEG distortion due to recompression forced by social media. The proposed
scheme relaxes this limitation.
\par
An evaluation of the proposed encryption scheme in terms of security and
compression showed that it enhances security against ciphertext-only attacks and that it is effective for EtC systems in terms of image quality.
\par

The rest of this paper is organized as follows. Section\,\ref{sec:preparation}
provides a review of conventional encryption schemes used in EtC systems.
Section\,\ref{sec:proposed} presents the proposed grayscale-based encryption
and its security enhancement. Extensive
experimental results including robustness
against jigsaw puzzle solver attacks are given
in Section\,\ref{sec:evaluation}.
Finally, Section\,\ref{sec:conclusion} concludes this paper.

\begin{figure}[t]
\centering
\includegraphics[width =7.6cm]{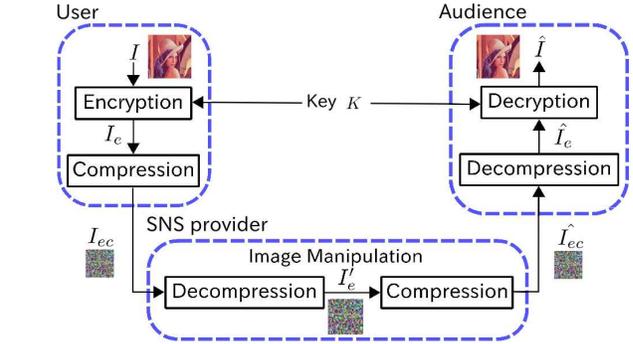}
\caption{EtC system}
\label{fig:etc}
\end{figure}

\begin{figure}[t]
\centering
\includegraphics[width =8.6cm]{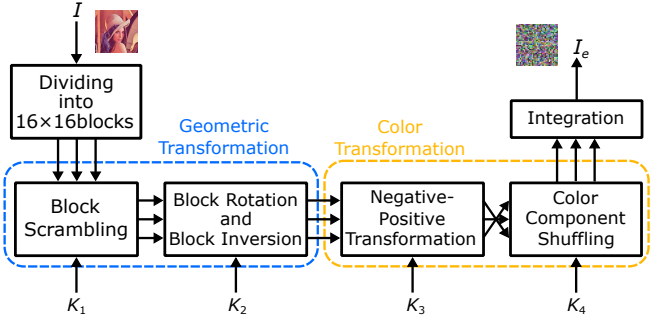}
\caption{Conventional block scrambling-based image encryption}
\label{fig:step}
\end{figure}

\begin{figure}[!t]
\centering
\hspace{2mm}\subfloat[Block rotation]{\includegraphics[width=3cm]{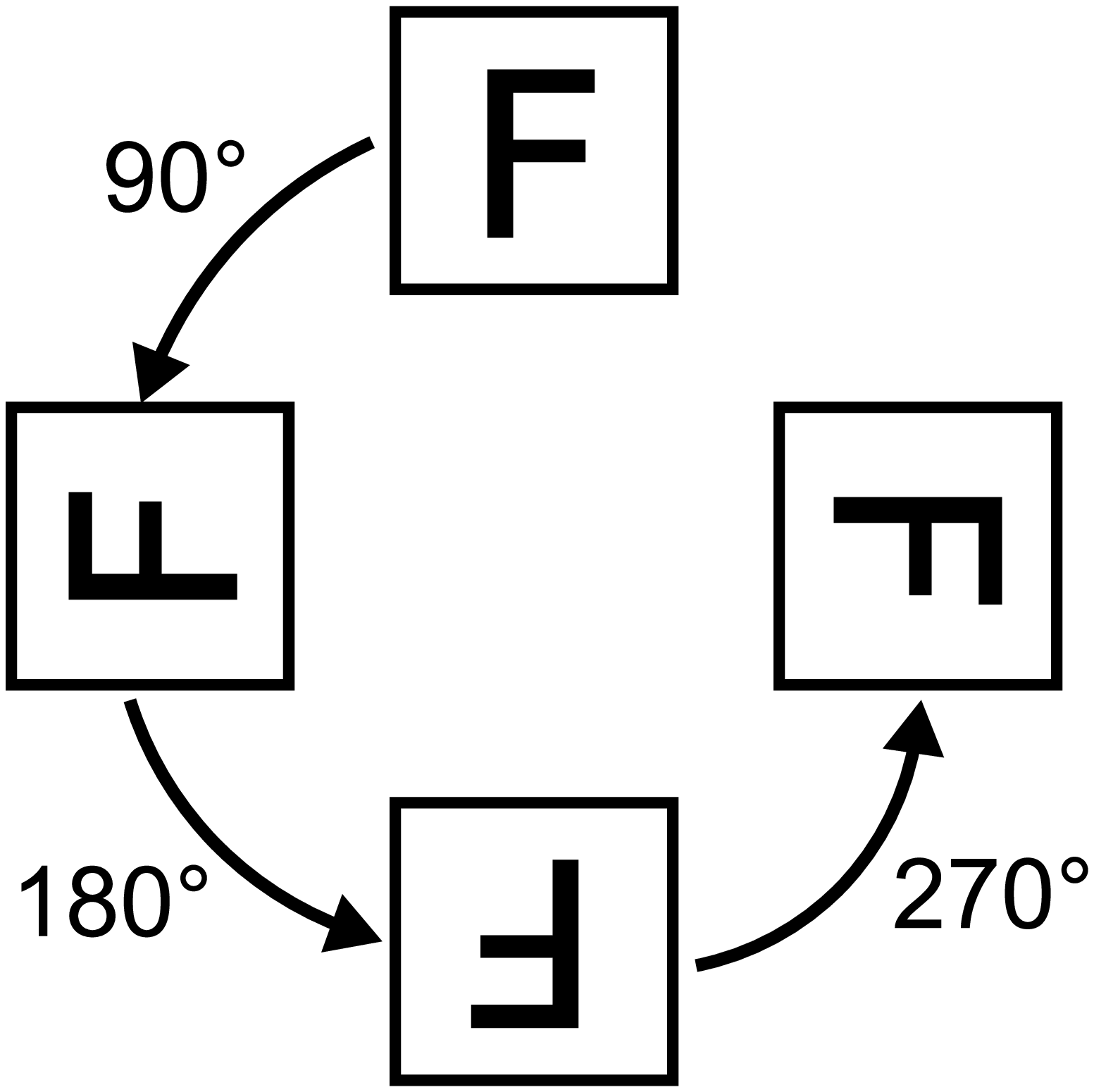}
}
\hfil
\subfloat[Block inversion]{\includegraphics[clip, width=3cm]{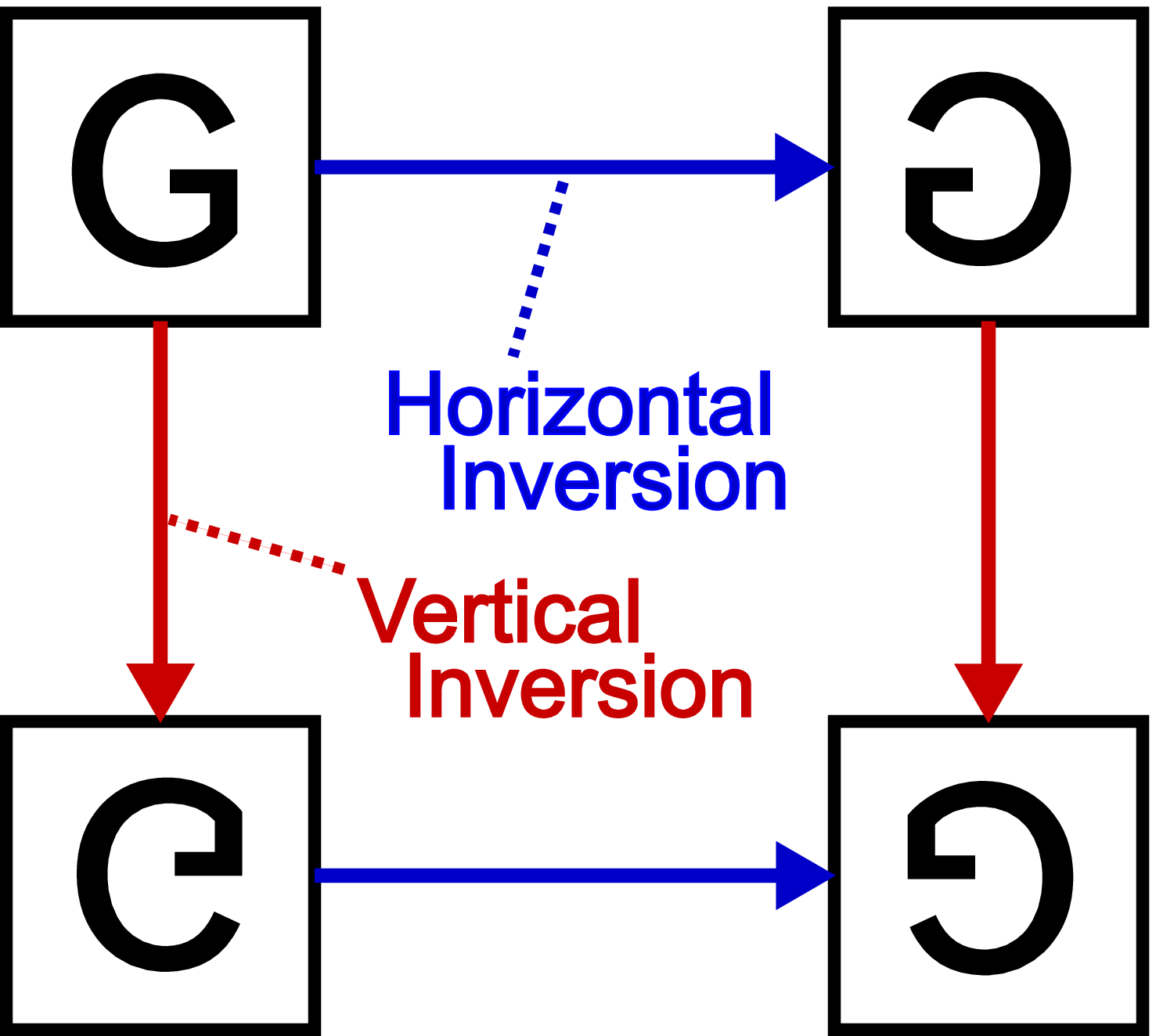}
}
\caption{Block rotation and inversion}
 \label{fig:rotinv}
\end{figure}

\begin{figure}[t]
\centering
\captionsetup[subfigure]{justification=centering}
\subfloat[Original image\newline($X \times Y$ = $384 \times 512$)]{
\includegraphics[clip, height=4.5cm]{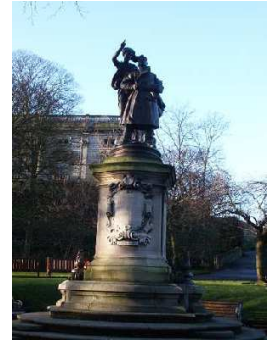}
\label{fig:ori}
}
\hspace{1cm}
\subfloat[Encrypted image\cite{kurihara2015encryption,KURIHARA2015} \newline($B_{x}\!=\!B_{y}\!=\!16, n\!=\!768$)]{
\includegraphics[clip, height=4.5cm]{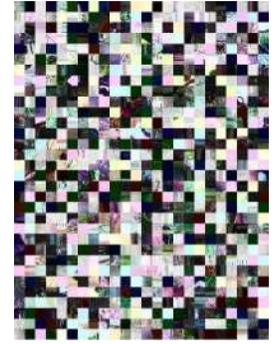}
\label{fig:enc}
}
\newline
\subfloat[Encrypted grayscale-based image using proposed scheme\newline(Rectangular, $B_{x}=B_{y}=8, n=9216$)]{
\includegraphics[clip, height=3.5cm]{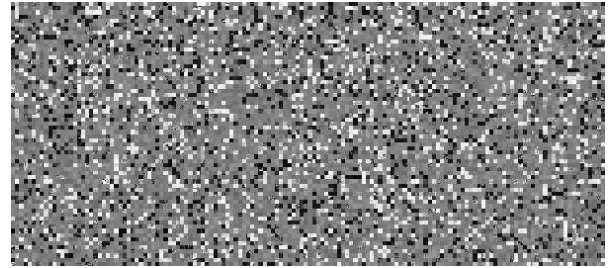}
\label{fig:enc2}
}
\newline
\subfloat[Encrypted graycale-based image using proposed scheme\newline(Square, $B_{x}=B_{y}=8, n=9216$)]{
\includegraphics[clip, height=4.9cm]{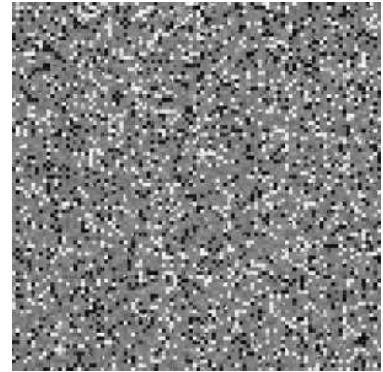}
\label{fig:enc_sq}
}
\caption{Encrypted images with the conventional and proposed scheme}
\label{fig:eximages}
\end{figure}

\section{Preparation}
\label{sec:preparation}
In this section, after the conventional block scrambling-based image encryption scheme\cite{kurihara2015encryption,KURIHARA2015,KuriharaBMSB,Kuri_2017} is summarized, the security of the scheme against brute-force and jigsaw puzzle solver attacks as ciphertext-only attacks (COA) is addressed.

\subsection{Block Scrambling-Based Image Encryption}
\label{blocksc}
A block scrambling-based image encryption scheme was proposed for EtC
systems\cite{watanabe2015encryption,kurihara2015encryption,KURIHARA2015,KuriharaBMSB,Kuri_2017}, in which a user wants to securely transmit image $I$ to an audience, via a Social Networking Service (SNS) provider, as illustrated in Fig.\,\ref{fig:etc}.
Since the user does not give the secret key $K$ to the SNS provider, the privacy
of image to be shared is under control of the user even when the SNS provider
recompresses image $I$. Therefore, the user can ensure privacy
by him/herself.
In comparison, in CtE systems, the user has to disclose unencrypted images
to recompress them.
\par
In the
scheme\cite{kurihara2015encryption,KURIHARA2015,KuriharaBMSB,Kuri_2017},
an image with $X \times Y$ pixels is first divided into non-overlapped blocks
with $B_x \times B_y$; then four block scrambling-based processing steps
are applied to the divided image.
Figure\,\ref{fig:step} illustrates the procedure of the scheme with $B_{x}
= B_{y} =16$.
In this paper, $B_{x} = B_{y} =16$ is used as well as in \cite{kurihara2015encryption,KURIHARA2015}.
The procedure for performing image encryption to generate an encrypted image
$I_e$ is given as follows.
\begin{itemize}
\item[Step 1:] Divide image with $X \times Y$ pixels into blocks, each with $B_x \times B_y$ pixels, and permute randomly the divided blocks using a random integer generated by a secret key $K_1$, where $K_1$ is commonly used for all color components.
\item[Step 2:] Rotate and invert randomly each block (see
Fig.\,\ref{fig:rotinv}) by using a random integer generated by a key $K_2$,
where $K_2$ is commonly used for all color components as well.
\item[Step 3:] Apply negative-positive transformation to each block by
using a random binary integer generated by a key $K_3$, where $K_3$ is commonly used for all color components. In this step, a transformed pixel value in the $i$th block $B_i$, $p'$, is computed using
\begin{equation}
p'=
\left\{
\begin{array}{ll}
p & (r(i)=0) \\
p \oplus (2^L-1) & (r(i)=1)
\end{array} ,
\right.
\end{equation}
where $r(i)$ is a random binary integer generated by $K_3$, and $p \in B_i$ is
the pixel value of the original image with $L$ bit per pixel. In this paper, the
value of occurence probability $P(r(i))=0.5$ has been used to invert bits
randomly.
\item[Step 4:] Shuffle three color components in each block by using an
integer randomly selected from six integers by a key $K_4$.
\end{itemize}
An example of an encrypted image ($B_{x}=B_{y}=16$) is shown in Fig.\,\ref{fig:eximages}\subref{fig:enc}; Fig.\,\ref{fig:eximages}(a) shows the original one. In this paper, we focus on block scrambling-based image encryption for the following reasons.
\begin{itemize}
\item[(a)]The encrypted images are compatible with the JPEG standard.
\item[(b)]The compression efficiency for the encrypted images is almost the same as that for the original ones under the JPEG standard.
\item[(c)]Robustness against various attacks has been demonstrated\cite{CHUMAN2017ICASSP,CHUMAN2017ICME}.
\end{itemize}
\par
The conventional encryption used in EtC systems has a limitation on block
size, i.e. $B_x=B_y=16$, to avoid the effect of color sub-sampling. If
$B_x=B_y=8$ is chosen as a block size, the compression performance decreases
and some block distortion is generated in decompressed images. In the JPEG
standard, when the color sub-sampling is applied to the chroma components
$C_b$ and $C_r$ of an color image in an encoder, the sub-sampled chroma
components are interpolated to reconstruct the same resolution as that of the original
image in a decoder. If 4:2:0 color sub-sampling is applied to encrypted images, 
each $8 \times 8$-block in the sub-sampled chroma components
consists of four $4 \times 4$-blocks from different $8 \times 8$-blocks, which
have generally a low correlation among the blocks. Therefore, the compression performance of the encrypted
images decreases, and moreover block distortion is generated due to the
interpolation of the sub-sampled chroma components with discontinuous values.

\subsection{Security against Ciphertext-only Attacks}
\subsubsection{Security}

Security mostly refers to protection from adversarial forces. The proposed
encryption scheme aims to protect visual information of images that allow us to
identify an individual, a time and the location for taking a photograph.
Untrusted providers and unauthorized users are assumed as the adversary. In this
paper, we consider security against not only brute-force attack, but also
jigsaw puzzle solver attacks as ciphertext-only attacks. The robustness against
jigsaw puzzle solver attacks is objectively evaluated in terms of the ration of
correct block positions and its computational cost.
\subsubsection{Brute-force Attack}
If an image with $X \times Y$ pixels is divided into blocks with $B_x \times B_y$ pixels, the number of blocks $n$ is given by
\begin{equation}
\label{eq:blocknum}
n = \lfloor \frac{X}{B_x} \rfloor \times \lfloor \frac{Y}{B_y} \rfloor ,
\end{equation}
where $\lfloor \cdot \rfloor$ is a function that rounds down to the nearest
integer. The four block scrambling-based processing steps are then applied
to the divided image.
\par
The key space of the block scrambling (Step 1) $N_S(n)$, which is the number of permutation of $n$ blocks, is given by
\begin{equation}
N_{S}(n) = { }_{n} P { }_{n} = n!.
\end{equation}

Similarly, the key spaces of other encryption steps are given as
\begin{equation}
N_{R}(n) = 4^{n},\;N_{I}(n) = 4^{n},\;N_{R\&I}(n) = 8^{n}\;
\end{equation}
\begin{equation}
N_{N}(n) = 2^{n}, \;
N_{C}(n) = \bigl( { }_3 P { }_3 \bigr)^{n} = 6^{n}
\end{equation}
where $N_{R}(n)$ and $N_{I}(n)$ are the key spaces of the block rotation and
block inversion, and $N_{R\&I}(n)$ is the key space of the encryption combining
them (Step 2). Note that $N_{R\&I}$ is the key space considering the collision
between block rotation and inversion. Namely, rotating pieces 180 degrees is the
same operation as inverting them horizontally and vertically. $N_N(n)$ and
$N_C(n)$ are the key spaces of the negative-positive transformation (Step3) and
color component shuffling (Step 4) respectively. Consequently, the key space of
images encrypted by using all the proposed encryption steps, $N_A(n)$, is
represented by
\begin{IEEEeqnarray}{ccc}
N_A(n) & = N_S(n) \cdot N_{R\&I}(n)\cdot N_N(n) \cdot N_C(n) \\
& = {n}! \cdot 8^{n} \cdot 2^{n} \cdot 6^{n}\nonumber.
\end{IEEEeqnarray}
The key space of the scheme is generally large enough against brute-force attacks such as ciphertext-only attacks\cite{KURIHARA2015}. However, by regarding the blocks of an encrypted image as pieces of a jigsaw puzzle, the attack can be regarded as a jigsaw puzzle solver attack.

\begin{figure}[!t]
\captionsetup[subfigure]{justification=centering}
\centering
\subfloat[Original image
\newline($X \times Y$ = $224 \times 160$)]
{\includegraphics[height=2.0cm]{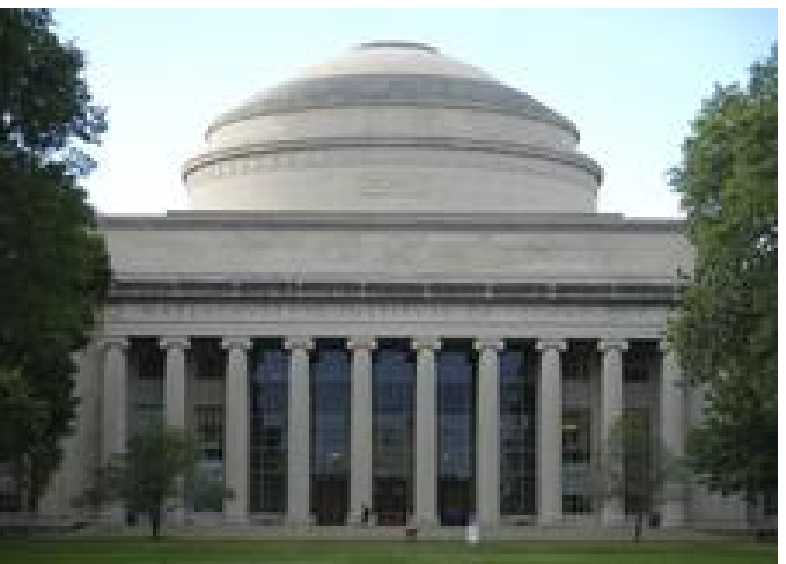}
\label{fig:jig_orig}}
\subfloat[Encrypted image
\newline(Conventional scheme,
\newline $\!B_{x}\!=\!B_{y}\!=\!16,n=\!140$)]
{\includegraphics[height=2.0cm]{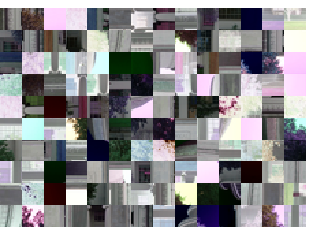}
\label{fig:jig_enc16}}
\subfloat[Encrypted image
\newline(Conventional scheme,
\newline $\!B_{x}\!=\!B_{y}\!=\!8,n=\!560$)]
{\includegraphics[height=2.0cm]{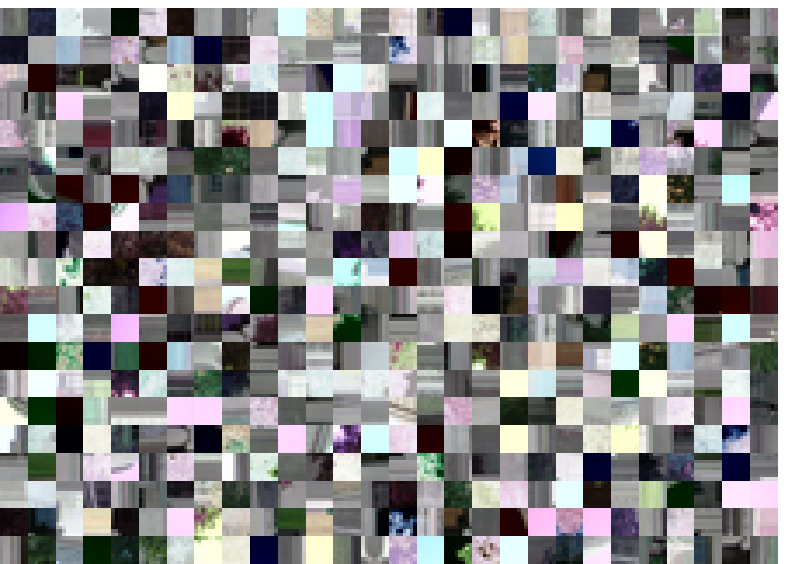}
\label{fig:jig_enc8}}
\\
\captionsetup[subfigure]{justification=centering}
\subfloat[Assembled image
\newline ($B_{x}=B_{y}=16$)]
{\includegraphics[height=2.0cm]{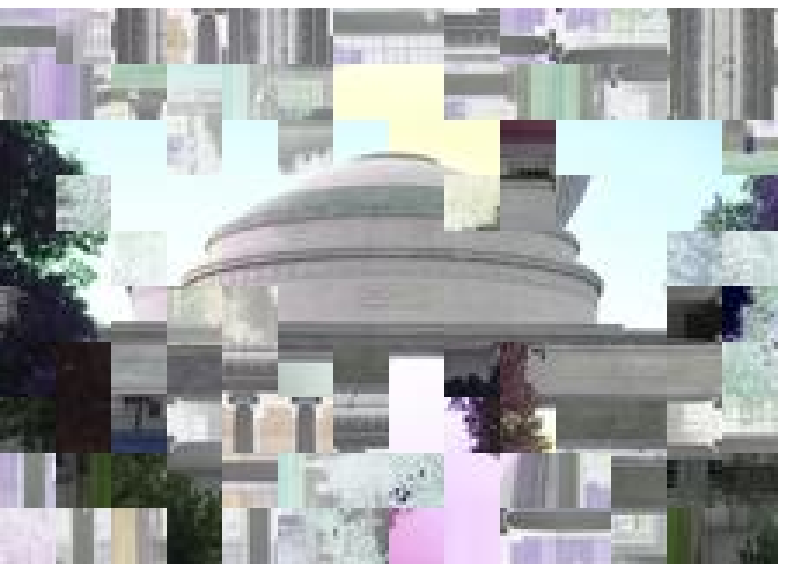}
\label{fig:jig_ass16}}
\subfloat[Assembled image
\newline ($B_{x}=B_{y}=8$)]
{\includegraphics[height=2.0cm]{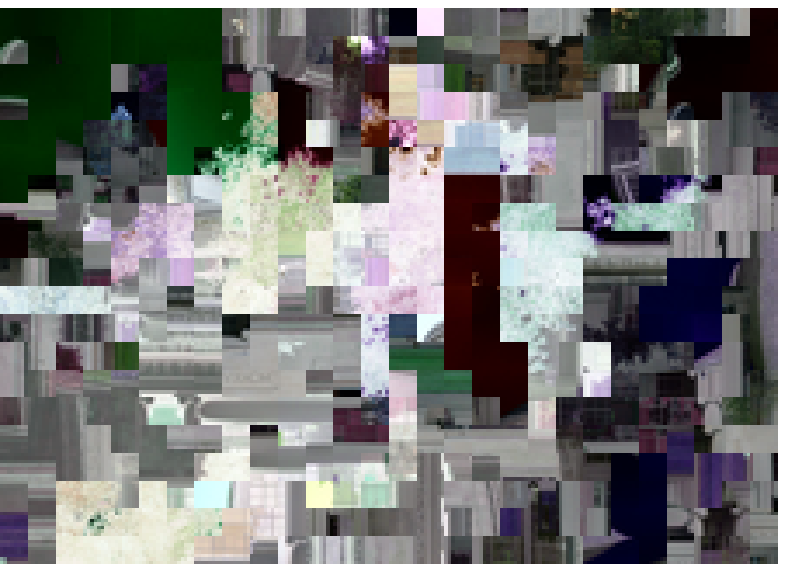}
\label{fig:jig_ass8}}
\caption{Assembled images by using the extended jigsaw puzzle solver\cite{CHUMAN2017IEICE}}
\label{fig:jigsaw_pre}
\end{figure}

\subsubsection{Jigsaw Puzzle Solver Attack}
To assemble encrypted images including inverted, negative-positive transformed and color component shuffled blocks, extended jigsaw puzzle solvers for block scrambling-based image encryption have been proposed\cite{CHUMAN2017ICASSP,CHUMAN2017ICME}.
It has been shown that assembling jigsaw puzzles becomes difficult when the encrypted images are satisfied with under three conditions\cite{CHUMAN2017ICASSP,CHUMAN2017ICME,CHUMAN2017IWSDA,CHUMAN2017IEICE}.
\begin{itemize}
\item[(a)]Number of blocks is large.
\item[(b)]Block size is small.
\item[(c)]Encrypted images include JPEG distortion.
\end{itemize}
Figures \ref{fig:jigsaw_pre}\subref{fig:jig_ass16} and \subref{fig:jig_ass8}
are examples of images assembled from
Figs.\,\ref{fig:jigsaw_pre}\subref{fig:jig_enc16} and \subref{fig:jig_enc8}
respectively; Fig.\,\ref{fig:jigsaw_pre}\subref{fig:jig_orig} shows the original one.
Compared with Figs.\,\ref{fig:jigsaw_pre}(d) and (e), the difficulty of
assembling encrypted images strongly depends on the block size.
In addition, most conventional jigsaw puzzle
solvers also utilize color information to assemble puzzles. Thus, reducing
the number of color channels in each pixel makes assembling encrypted
images much more difficult.
The novel scheme in this paper has a higher security level than
that of the conventional scheme, because it provides a large number of blocks
and the small block size.
\par
Other attacking strategies such as known-plaintext attack (KPA) and
chosen-plaintext attack (CPA) should be considered for the security. The block
scrambling-based image encryption becomes robust against KPA through the
assigning of a different key to each image for the encryption. In
addition, the keys used for the encryption do not need to be disclosed because
the encryption scheme is not public key cryptography. Therefore, the encryption
can avoid the CPA, unlike public key cryptography.
\subsection{Summary of Image Encryption for EtC Systems}

The properties of the conventional encryption
scheme\cite{kurihara2015encryption,KURIHARA2015} that is only one
conventional scheme for EtC systems with the JPEG standard is summarized in
Table.\,\ref{classify}. The proposed one enables the use of a smaller block
size and a larger number of blocks, which enhances both invisibility and
security against several attacks. Furthermore, images encrypted by using the
proposed scheme include less color information due to the use of grayscale
images, which makes the EtC system more robust. The properties of the proposed
one allow us not only to enhance security against several attacks, but also to
avoid the effect of color sub-sampling.

\begin{table}[t]
\begin{center}
\begin{threeparttable}
\centering
\caption{Properties of block scrambling-based image encryption
schemes}
\label{classify}
\begin{tabular}{ccc}
Scheme                                                                          
& Conventional\cite{kurihara2015encryption,KURIHARA2015} & Proposed    \\ \hline Color channel                                                                             & RGB          & Grayscale   \\ \hline
\begin{tabular}[c]{@{}c@{}}Minimum block size\end{tabular}                              & $16 \times 16$ & $8 \times 8$  \\ \hline
\begin{tabular}[c]{@{}c@{}}Image size\end{tabular} & $X \times Y$       & $3
\times X \times Y$ \\ \hline Number of blocks                                                                          & $n$\tnote{*}          & $3n$        \\ \hline
\begin{tabular}[c]{@{}c@{}}Robustness against\\ jigsaw puzzle solver
attacks\end{tabular} & Robust       & More robust \\ \hline
\begin{tabular}[c]{@{}c@{}}Effect of color sub-sampling\end{tabular}            
& Affected  & Unaffected\\ \hline
\end{tabular}
\begin{tablenotes}
\item[*] Calculated from Eq.\,\ref{eq:blocknum}
\end{tablenotes}
\end{threeparttable}
\end{center}
\end{table}

\section{Proposed Method}
\label{sec:proposed}
In this section, the grayscale-based image encryption scheme, which has
higher security than the conventional one, is proposed.
\begin{figure}[t]
\centering
\includegraphics[width =8.6cm]{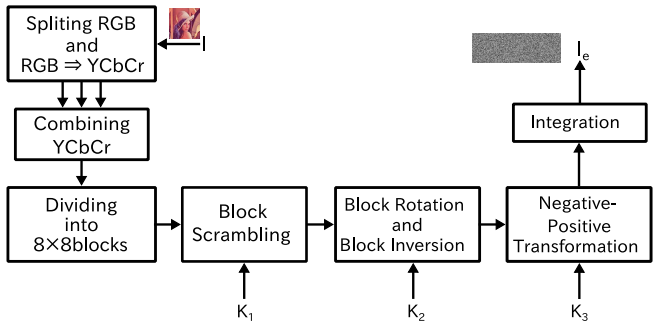}
\caption{Proposed grayscale-based image encryption}
\label{fig:step_pro}
\end{figure}
\subsection{Procedure of Proposed Image Encryption}
\label{new}
Although $B_{x}=B_{y}=16$ is used as the smallest block size in the conventional block scrambling-based image encryption to avoid the effect of color sub-sampling in JPEG compression, the proposed method enables us to use $B_{x}=B_{y}=8$ as a block size, which enhances robustness against ciphertext-only attacks.
Moreover, applying EtC systems with the proposed scheme to social media
performs better than with the conventional one as
described below.
\par
The procedure of the proposed scheme for an 8-bit RGB full-color image with $X \times Y$ pixels is given as follows (See Fig.\,\ref{fig:step_pro}):
\begin{itemize}
\item[1)]Split color image into three (RGB) channels.

\item[2)]
Considering JPEG compression efficiency, RGB components are transformed into
YCbCr color space by using the equation below, as in \cite{ITU}.
      \begin{IEEEeqnarray}{rrrrr}
      \!\!\!\!\!\!\!\!\!\!\!\!\!\!\!Y     = &  0.299*R\quad  &\!\!\! +0.587*G\quad  &\!\!\!  +0.114*B  &\!\!\! \label{eq:yeq}\\
      \!\!\!\!\!\!\!\!\!\!\!\!\!\!\!C_{b} = & -0.1687*R\quad &\!\!\! -0.3313*G\quad &\!\!\!  +0.5*B    &\quad\!\!\! +128  \\
      \!\!\!\!\!\!\!\!\!\!\!\!\!\!\!C_{r} = &  0.5*R\quad    &\!\!\! -0.4187*G\quad &\!\!\!  -0.0813*B &\quad\!\!\! +128
    \end{IEEEeqnarray}
\item[3)]Select $B_{x}=B_{y}=8$ as a block size.
\item[4)]Combine YCbCr channels into one grayscale-based image. A grayscale image with $3 \times X \times Y$ pixels is thereby generated.
\item[5)]Apply block scrambling, rotation, inversion, and negative-positive transformation by using Steps 1 to 3 in Sec.\,\ref{blocksc}.
\end{itemize}
An example of an encrypted image ($B_{x}=B_{y}=8$) is shown in Fig.\,\ref{fig:eximages}\subref{fig:enc2}, where YCbCr channels are combined horizontally.
As shown in Fig.\,\ref{fig:eximages}\subref{fig:enc_sq}, which is an example of an encrypted image combined YCbCr channels to become square, the way of combining YCbCr channels has some freedom.
\subsection{Compression of Grayscale-based Image}
In this paper, we focus on JPEG lossy compression,
although the JPEG standard supports both lossless and
lossy compressions, and the encryption schemes are
applicable to lossless compression methods as discussed
in\cite{Kuri_2017}. This is because most JPEG compression
applications, especially SNS providers and Cloud Photo
Storage Services (CPSS), use lossy compression, and lossless compression does
not generate any distortion.
\par
Whereas RGB images are generally compressed by using two quantization tables,
namely for luminance ($Y$) and for chrominance ($C_b$, $C_r$) respectively,
grayscale images are compressed with only one quantization table in the JPEG standard.
Images encrypted with the proposed scheme are grayscale-based, so one
quantization table is used.
The quantization table that is selected for the proposed scheme affects the compression performance.
In the experiments, the relationship between quantization tables and compression performances is discussed.
\subsection{Decompression and Decryption of Encrypted Images}

In order to reconstruct images from encrypted ones, JPEG images downloaded from
a provider are decompressed by a JPEG decoder, and then the decryption
process is applied to the decompressed images.
\par
In the JPEG decompression, the color interpolation is
not performed over images encrypted by the proposed method even when $8 \times
8$ is used as a block size, because the encrypted images contain no chroma
component. In contrast, the conventional scheme suffers from the block
distortion due to the interpolation of chroma components. \par
\begin{figure}[t]
\centering
\includegraphics[width =8.6cm]{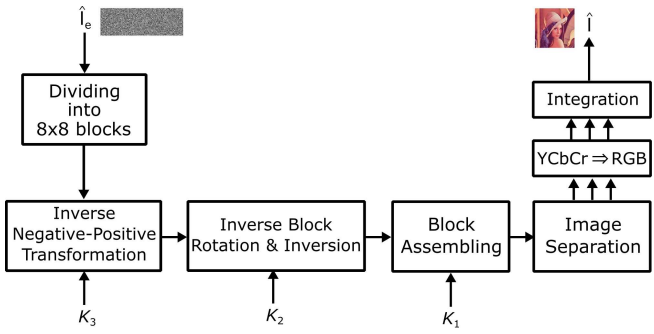}
\caption{Decryption process for grayscale-based image encryption}
\label{fig:step_pro_de}
\end{figure}

 After the
decompression process, six decryption steps are carried out using the corresponding secret key $K$ as follows (See Fig.\,\ref{fig:step_pro_de}).

\begin{itemize}
\item[1)]Divide the encrypted image into blocks, each with $B_x \times
B_y$ and apply inverse negative-positive transformation to each block with key
$K_3$.
\item[2)]Inversely rotate and invert each block with key
$K_2$.
\item[3)]Assemble blocks with key $K_1$.
\item[4)]Separate the grayscale-based image into three channels, $Y$,
$C_b$, and $C_r$, each with $M \times N$ pixels.
\item[5)]Transform the three channels to RGB color channels.
\item[6)]Combine the RGB channels to generate the decrypted
image.
\end{itemize}
\par

Note that the compression performance of the proposed scheme decreases when
other block sizes are used, such as $4 \times 4$ and $10 \times 10$, due to 
discontinuous pixel values in each block as well as for the conventional scheme,
although the color interpolation is not carried out. 

\subsection{Enhanced Security}
The proposed scheme has superior features compared to the conventional encryption one in terms of the block size and number of blocks. Furthermore, the encrypted images generated by the proposed scheme are grayscale ones. These features enhance both invisibility and robustness against brute-force and jigsaw puzzle solver attacks, as described below.

\subsubsection{Small Block Size}
Since chroma components $C_b$ and $C_r$ are sub-sampled in JPEG compression, a
color image must be commonly split into 16 $\times$ 16 blocks as an minimum coded unit (MCU) in order to make 8$\times$8 blocks. Therefore, when the block scrambling-based image encryption is applied to color images, the smallest block size is $B_{x}=B_{y}=16$ to avoid the influence of color sub-sampling. In contrast, when grayscale images are compressed using the JPEG standard, color sub-sampling is not carried out. Therefore, $B_{x}=B_{y}=8$ is selected as the smallest block size for grayscale images in the JPEG standard.
\par
The encrypted images generated by the proposed scheme are grayscale-based ones. Therefore, $B_{x}=B_{y}=8$ can be selected for the encryption, as shown in Fig.\,\ref{fig:eximages}\subref{fig:enc2} and \subref{fig:enc_sq}, even if the encrypted images are compressed using the JPEG standard.

\subsubsection{Large Number of Blocks }
When $B_{x}=B_{y}=8$ is selected, the number of blocks increases fourfold, $B_{x}=B_{y}=16$, from Eq.\,(\ref{eq:blocknum}). Moreover, the proposed scheme generates encrypted images with $3 \times X \times Y$ pixels from an original image with $X \times Y$ pixels. As a result, the number of blocks is 12 times that with the conventional one, as shown in Figs.\,\ref{fig:eximages}(b) and (c).
\par
The running time to assemble encrypted images strongly depends on the
number of pieces when jigsaw puzzle solver attacks are utilized.
Furthermore, increasing the number of blocks makes assembling encrypted
images much more difficult.
Therefore, the proposed scheme enhances security.

\subsubsection{Less Color Information in Blocks}
The encrypted images include only one channel per pixel due to the use of grayscale-based images. This makes assembling jigsaw puzzles more difficult because almost all solvers utilize three color channels in each block to assemble puzzles correctly\cite{Gallagher_2012_CVPR,Son_2016_CVPR,Paikin_2015_CVPR,Andalo_2016}. Therefore, robustness against jigsaw puzzle solver attacks is enhanced.
\subsubsection{Key Space Expansion}
Using the proposed image encryption makes the key space of encrypted images larger than with the conventional scheme due to the increased number of blocks. The key space of the proposed scheme $N_{B}(n)$ is given by
\begin{IEEEeqnarray}{rrl}
\label{neweq}
N_B(n) & = N_{S}(3n) \cdot N_{R\&I}(3n) \cdot N_N(3n) \\
& = {3n}! \cdot 8^{3n} \cdot 2^{3n} \gg N_{A}(n)\nonumber ,
\end{IEEEeqnarray}
where $n$ is the number of blocks, calculated from an original image with $X \times Y$ pixels in accordance with Eq.\,(\ref{eq:blocknum}). Although the proposed scheme does not apply color component shuffling, the number of blocks is larger, as shown in Eqs.\,(\ref{eq:blocknum}) and (\ref{neweq}). Thus, the proposed scheme enhances robustness against brute-force attacks.
\begin{figure}[!t]
\captionsetup[subfigure]{justification=centering}
\centering
 \hspace{2mm}\subfloat[Decrypted image with block artifact (Conventional scheme, PSNR=31.4dB, sub-sampling ratio=4:2:0)]{\includegraphics[clip, width=3.9cm]{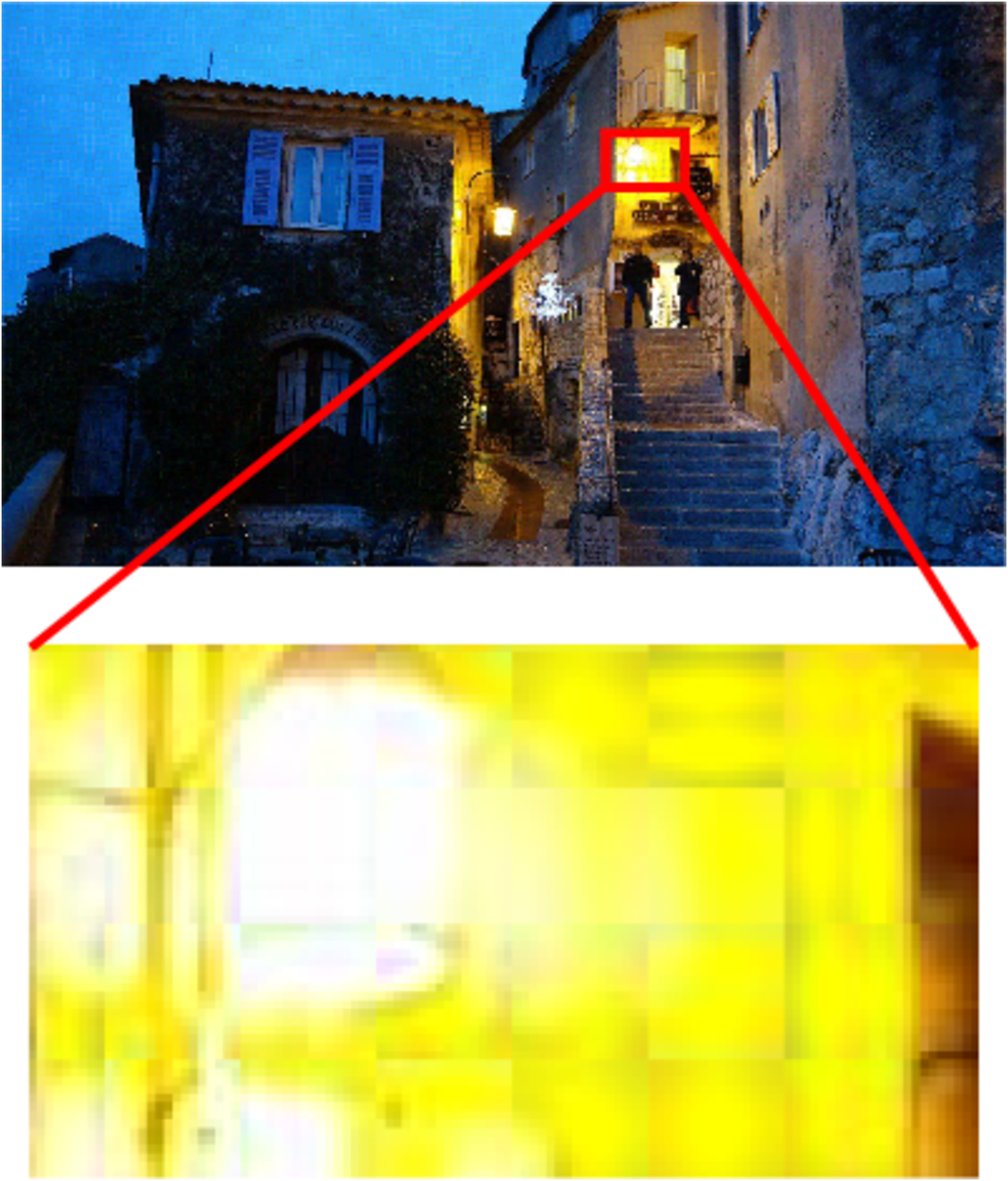}
\label{fig:noiseimage}}
\hfil
\subfloat[Decrypted image (Proposed scheme, PSNR=36.3dB)]{\includegraphics[clip, width=3.9cm]{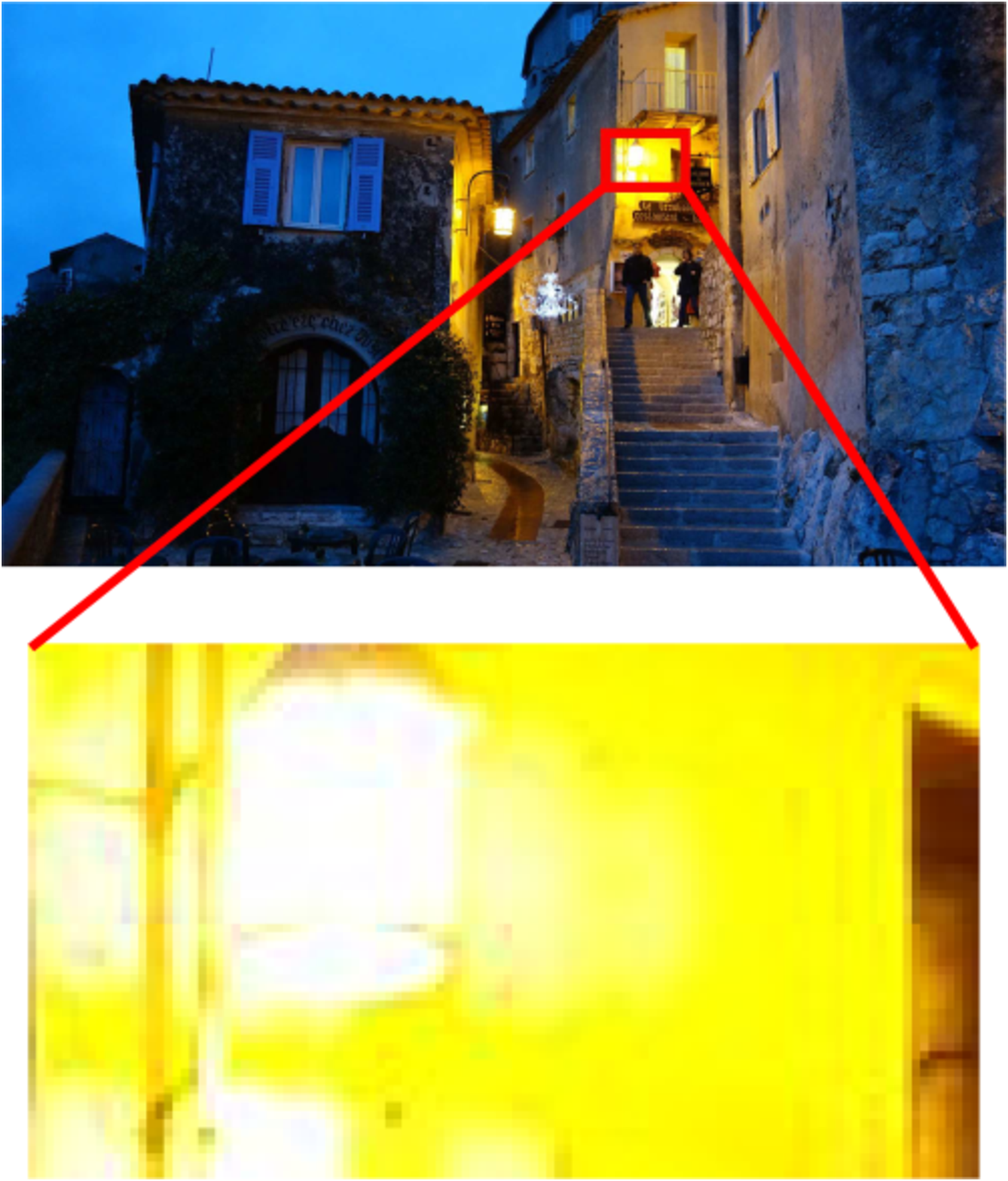}
\label{fig:label-C}}
\caption{Decrypted images downloaded from Facebook. The image in (a) includes some block distortion due to the effect of color sub-sampling.}
 \label{fig:oriencex}
\end{figure}

\begin{table*}[t]
\centering
\caption{Relationship between uploaded JPEG files and downloaded ones in terms
of sub-sampling ratios and the maximum resolutions. Providers do not resize
uploaded images when their resolutions are less than or equal to the maximum
resolutions, e.g. the maximum resolutions of Twitter and  Tumblr are $4096
\times 4096$ and $1280 \times 1280$, respectively.}
\begin{threeparttable}
\begin{tabular}{|c||c|c||c|c|}
\hline
\multirow{2}{*}{SNS provider} & \multicolumn{2}{c||}{Uploaded JPEG file} & \multicolumn{2}{c|}{Downloaded JPEG file} \\ \cline{2-5} 
 & \begin{tabular}[c]{@{}c@{}}Sub-sampling ratio\end{tabular} & $Q_{fu}$ & \begin{tabular}[c]{@{}c@{}}Sub-sampling  ratio\end{tabular} & $Q_{fd}$ \\ \hline
\multirow{6}{*}{Twitter (Up to 4096$\times$4096 pixels)} & \multirow{2}{*}{4:4:4} & low & \multicolumn{2}{c|}{No recompression} \\ \cline{3-5} 
 &  & high & 4:2:0 & 85 \\ \cline{2-5} 
 & \multirow{2}{*}{4:2:0} & 1,2,\ldots84 & \multicolumn{2}{c|}{No recompression} \\ \cline{3-5} 
 &  & 85,86,\ldots100 & 4:2:0 & 85 \\ \cline{2-5}
 & \multirow{2}{*}{(Grayscale)} & 1,2,\ldots84 & \multicolumn{2}{c|}{No recompression} \\ \cline{3-5} 
 &  & 85,86,\ldots100 & (Grayscale) & 85\\ \hline\hline
\multirow{3}{*}{\begin{tabular}[c]{@{}c@{}}Facebook (HQ, Up to 2048$\times$2048 pixels)\\ Facebook (LQ, Up to 960$\times$960 pixels)\end{tabular}} & 4:4:4 & \multirow{3}{*}{1,2,\ldots100} & \multirow{2}{*}{4:2:0} & \multirow{3}{*}{71,72,\ldots85} \\ \cline{2-2}
 & 4:2:0 &  &  &  \\ \cline{2-2}\cline{4-4} 
  & (Grayscale) &  & (Grayscale) &  \\ \hline\hline 
\multirow{5}{*}{\begin{tabular}[c]{@{}c@{}}Tumblr (Up to 1280$\times$1280 pixels)\\ Google+\\ Flickr\end{tabular}} & \multirow{2}{*}{4:4:4} &  & \multicolumn{2}{c|}{\multirow{5}{*}{No recompression}} \\
 &  & \multirow{3}{*}{1,2,\ldots100} & \multicolumn{2}{c|}{} \\ \cline{2-2}
 & \multirow{2}{*}{4:2:0} &  & \multicolumn{2}{c|}{} \\
 &  &  & \multicolumn{2}{c|}{} \\ \cline{2-2}
 & (Grayscale) &  & \multicolumn{2}{c|}{} \\ \hline
\end{tabular}

\end{threeparttable}
\label{tb:change_sampling}
\end{table*}

\subsection{Application to Social Media}
Figure\,\ref{fig:etc} illustrates the application of the EtC system to social media such as Facebook and Twitter, where the user wants to securely transmit image $I$ to an audience via social media. Since the user does not give the secret key $K$ to social media providers, the privacy of the image to be shared is under the user's control, even if the social media provider decompresses the image. This means that, if an encrypted image saved on the provider's server is leaked, third parties and general audiences cannot see the images visually unless they have the key.
\par
Table\,\ref{tb:change_sampling} summarizes the relationship between uploaded
and downloaded JPEG images in terms of sub-sampling ratios and JPEG
quality factor $Q_{f}$.
The previous work\cite{CHUMAN2017APSIPA} discussed the relationship for color images, but Table\,\ref{tb:change_sampling} includes the relationship for grayscale images.
$Q_{fu}$ and $Q_{fd}$ indicate the quality factor of uploaded and downloaded
images, respectively.
Although it has been confirmed that the EtC system with the conventional scheme is applicable to social media, EtC systems using the proposed scheme have superior features to the conventional one as shown below.
\subsubsection{Color Sub-sampling}
\label{subsampling}
As indicated in Table \ref{tb:change_sampling}, SNS providers manipulate uploaded images by changing the sub-sampling ratio and $Q_{fu}$.
Thus, we have to consider the effect of sub-sampling ratios, when encrypted images are upload to SNS providers that recompress uploaded images.
JPEG images with 4:2:0 sub-sampling ratio are interpolated to increase
the spatial resolution for chroma components in the decoding process.
This interpolation processing is carried out by using the relationship among blocks.
Therefore, encrypted images with 4:2:0 sub-sampling ratio are affected by
this interpolation.
As shown in Fig.\,\ref{fig:oriencex}(a), the decrypted image includes
block distortion due to the interpolation on Facebook.
In comparison, the interpolation in the decoding process does not affect
encrypted images with the proposed scheme, so encrypted images can avoid
the distortion as shown in Fig.\,\ref{fig:oriencex}(b).
Thus, EtC systems with the proposed scheme can avoid the effects of the interpolation, which is carried out on social media.
\subsubsection{Recompression}
Furthermore, images encrypted with the proposed scheme perform better
in terms of image quality.
The quality of images downloaded from social media is generally
degraded due to recompression forced by SNS providers, when the operation
of color sub-sampling is carried out.
Images encrypted by the proposed scheme always avoid this operation, which enable users to transmit images of higher quality than using the conventional scheme.
As well as SNS providers, EtC systems with the proposed scheme are applicable to Cloud Photo Services like iCloud and Google Photos. 
\subsubsection{Resizing}
To apply EtC systems to social media, the resolution of encrypted images needs
to be smaller than the maximum resolutions that each provider decides on as a
resizing condition\cite{CHUMAN2017APSIPA}.
For example, Twitter does not resize uploaded images with the
resolution of up to $4096 \times 4096$ pixels, as shown in
Table\,\ref{tb:change_sampling}. Resizing the resolution of encrypted images
makes the block size of the encrypted images smaller, although the JPEG
compression is still carried out based on the size of $8 \times 8$. As a
result, each $8 \times 8$-block in resized images includes pixels from
originally different blocks, so the compression performance decreases and block distortion
is generated in the decrypted images due to the discontinuity among pixels.
Thus, we have to upload encrypted images within the maximum resolution as indicated in Table\,\ref{tb:change_sampling}.
In this paper, the use of the image resolution that is not forcedly resized by SNS providers is assumed as well as the conventional EtC systems\cite{CHUMAN2017APSIPA}.
\section{Evaluation}
\label{sec:evaluation}
In this section, we evaluated the effectiveness of the proposed scheme in a
number of simulations.
First, we evaluated the compression performance of images encrypted by the proposed scheme.
Next, encrypted images were uploaded to SNS providers and then downloaded
from the providers to confirm robustness against image manipulation on
social media.
Finally, the security enhancement of the EtC systems against jigsaw puzzle
solver attacks\cite{CHUMAN2017IEICE} is shown from the aspect of the
difficulty of assembling encrypted images.
\subsection{Compression Performance}
\label{single_compress}
To determine the compression performance of the proposed scheme, a lot of
images encrypted by the conventional\cite{KURIHARA2015} and proposed scheme were compressed and decompressed.
Then, we calculated peak signal-to-noise ratio (PSNR) between original images
and images encrypted with the proposed method. Uncompressed Color Image
Database(UCID) dataset, which contains 1338 color images with the sizes of 512$\times$384 or 384$\times$512, was used for the evaluation.
The average PSNR values of all images per $Q_{f}$ in the dataset were used for the evaluation.
Software by the Independent JPEG Group's (IJG) \cite{JPEGLIB} was used for
encoding and decoding the images with following parameters.
\begin{itemize}
\item[$\cdot$]Sub-sampling ratio : 4:4:4, 4:2:0
\item[$\cdot$]Quantization table : IJG standard table
\end{itemize}
Non-encrypted images were also compressed with the same parameters.
As we mentioned before, grayscale-based images are generally compressed with
only one quantization table, so the luminance and chrominance tables
are used respectively for encoding images encrypted with the proposed
scheme.
In this paper, the quantization tables for luminance and chrominance designed by IJG were used.
\par
\begin{figure}[t]
\centering
\captionsetup[subfigure]{justification=centering}
\subfloat[Sub-sampling ratio : 4:2:0]{
\includegraphics[clip,width=8cm]{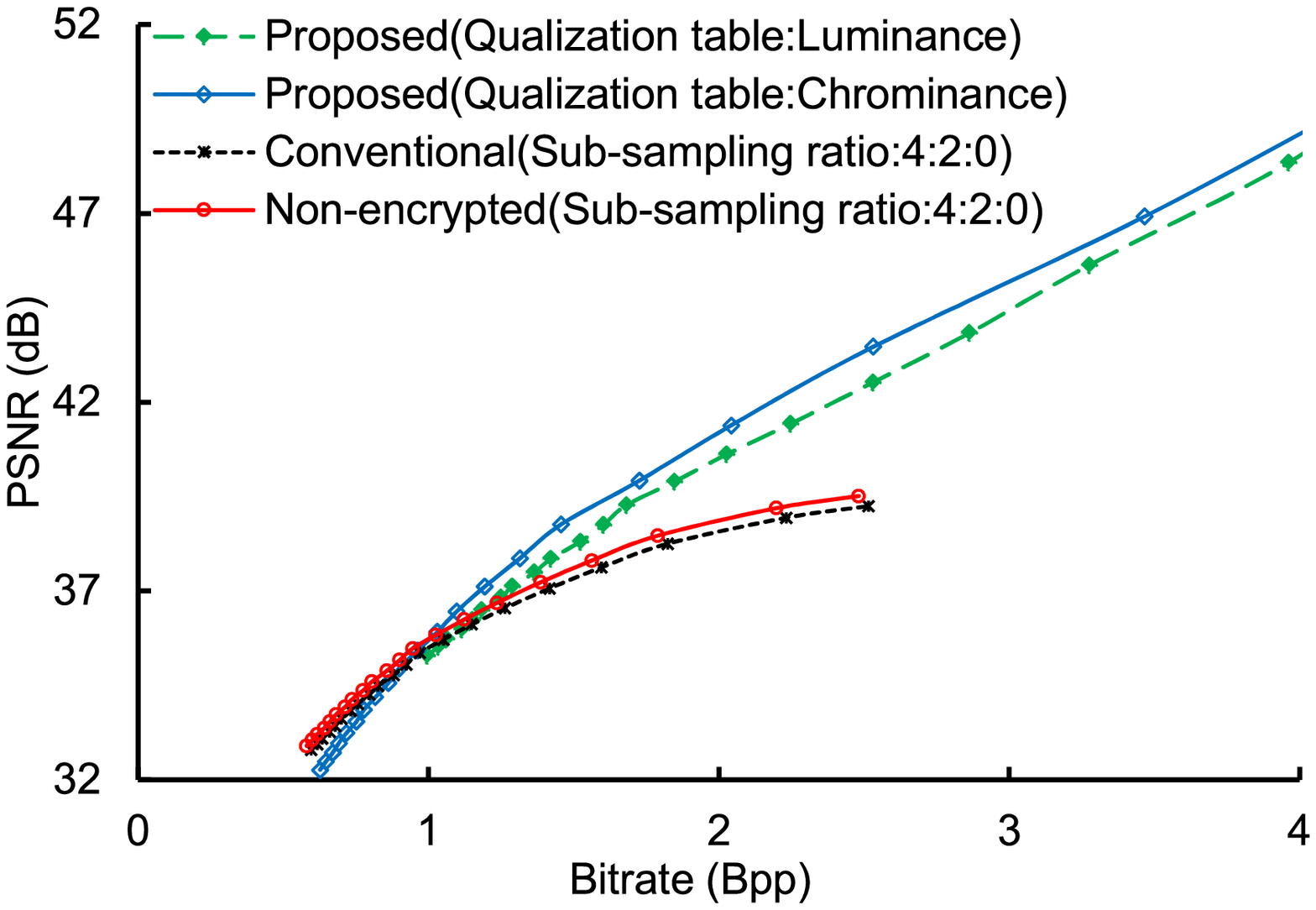}
\label{fig:420}
}
\newline
\subfloat[Sub-sampling ratio : 4:4:4]{
\includegraphics[clip, width=8cm]{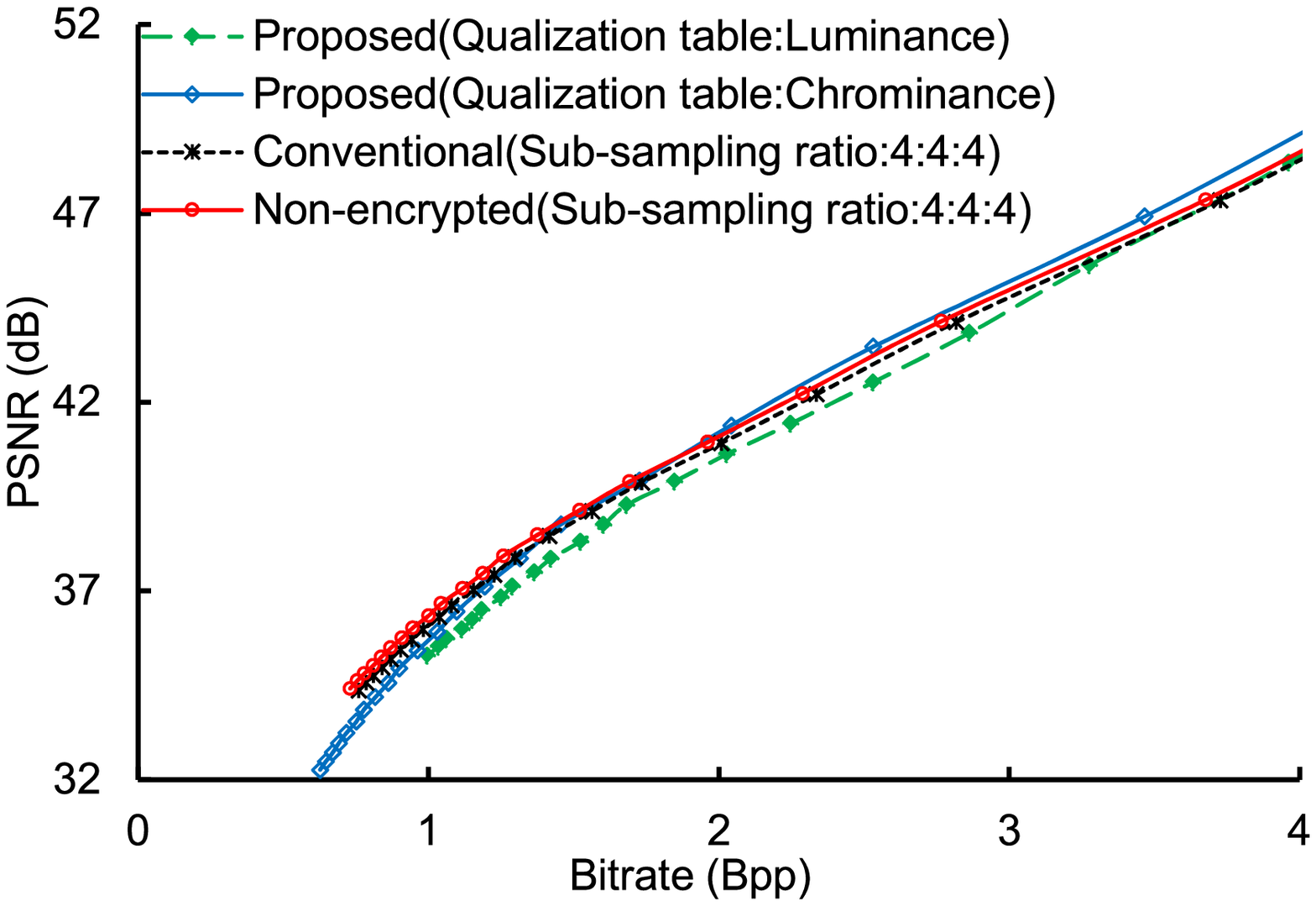}
\label{fig:444}
}
\caption{RD curves of original images and encrypted ones}
\label{fig:single_compression_ucid}
\end{figure}

Figure\,\ref{fig:single_compression_ucid} shows rate-distortion (RD) curves
of JPEG compressed images without any encryption and with encryption.
As the bitrate increased, the encrypted images with the proposed scheme had better compression performance than non-encrypted images compressed with 4:2:0 sub-sampling ratio as shown in Fig.\,\ref{fig:single_compression_ucid}\subref{fig:420}.
From the results, the operation of color sub-sampling affected the compression
performance as the bitrate increases.
In terms of the relationship between the proposed scheme with chrominance table
and with luminance table, the compression performance was very similar.
Fig.\,\ref{fig:single_compression_ucid}\subref{fig:444} shows the result
with 4:4:4 sub-sampling ratio.
The compression performance of the proposed scheme was almost the same as
the non-encrypted images compressed with 4:4:4 sub-sampling ratio.
Therefore, it was determined that images encrypted by the proposed scheme
have almost the same compression performance as both non-encrypted and conventional ones.
\par
\begin{figure}[!t]
\captionsetup[subfigure]{justification=centering}
\centering
\subfloat[Original image
\newline($X \times Y$ = $512 \times 512$)]
{\includegraphics[height=3cm]{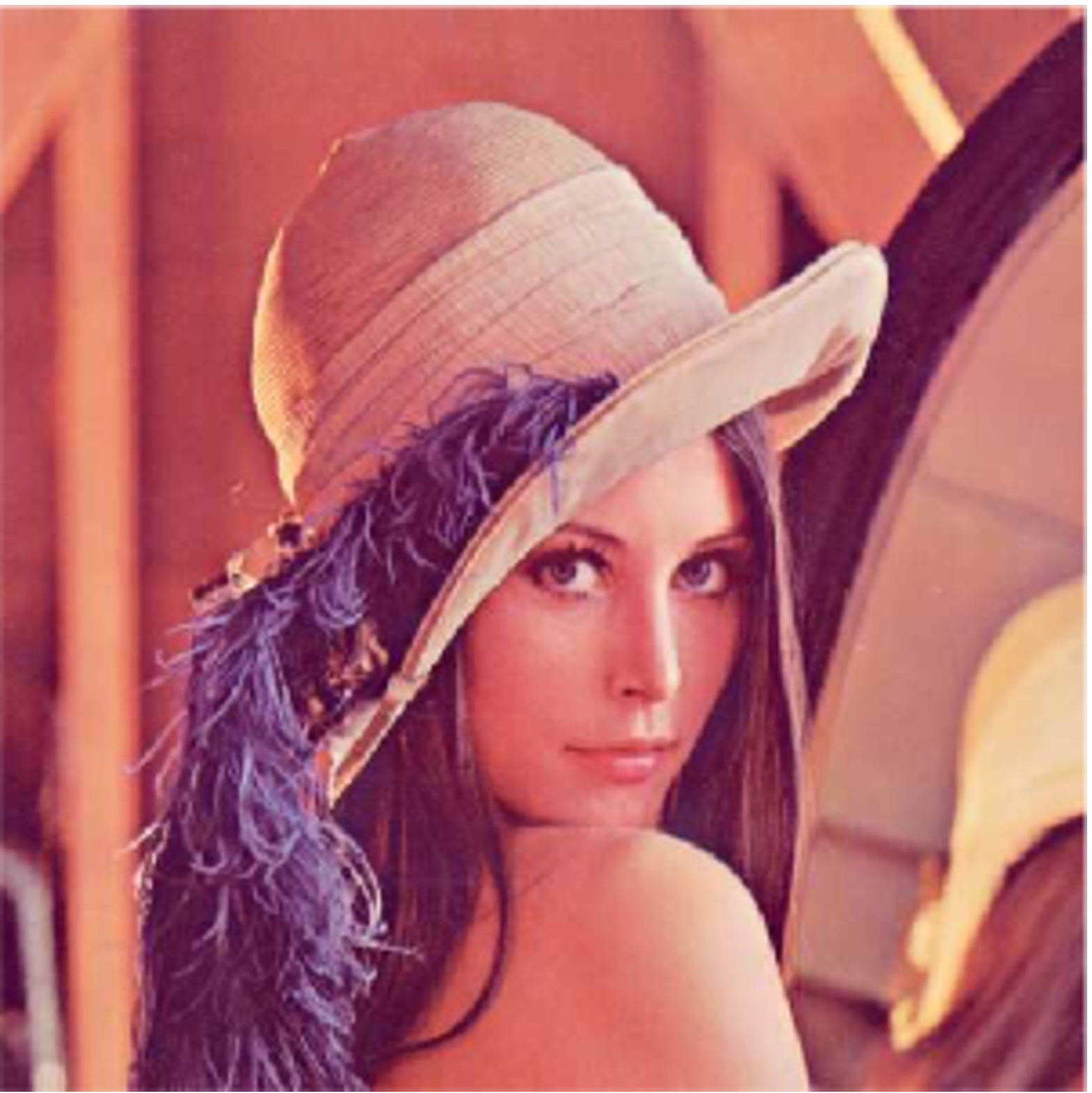}
\label{fig:orig_lena}}
\subfloat[Proposed scheme (PSNR=39.4dB)]
{\includegraphics[height=3cm]{./image/LENA.eps}
\label{fig:dec_lena_pro}}
\\
\subfloat[Encryption scheme in \cite{2016_Gaata} (PSNR=10.32dB, sub-sampling
ratio=4:2:0)]
{\includegraphics[height=3cm]{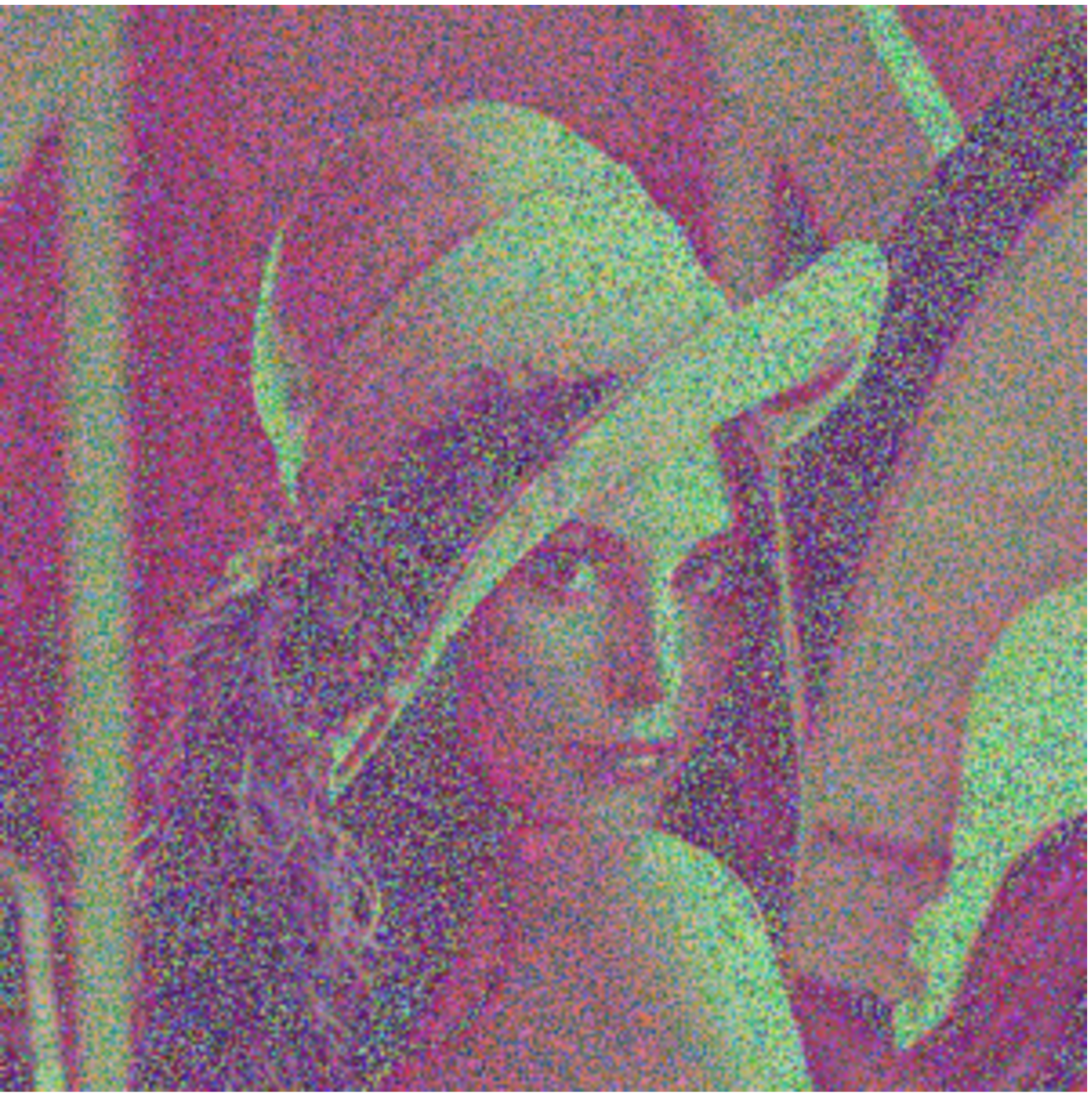}
\label{fig:dec_lena_gaata}}
\subfloat[Encryption scheme in \cite{2012_wu}]
{\includegraphics[height=3cm]{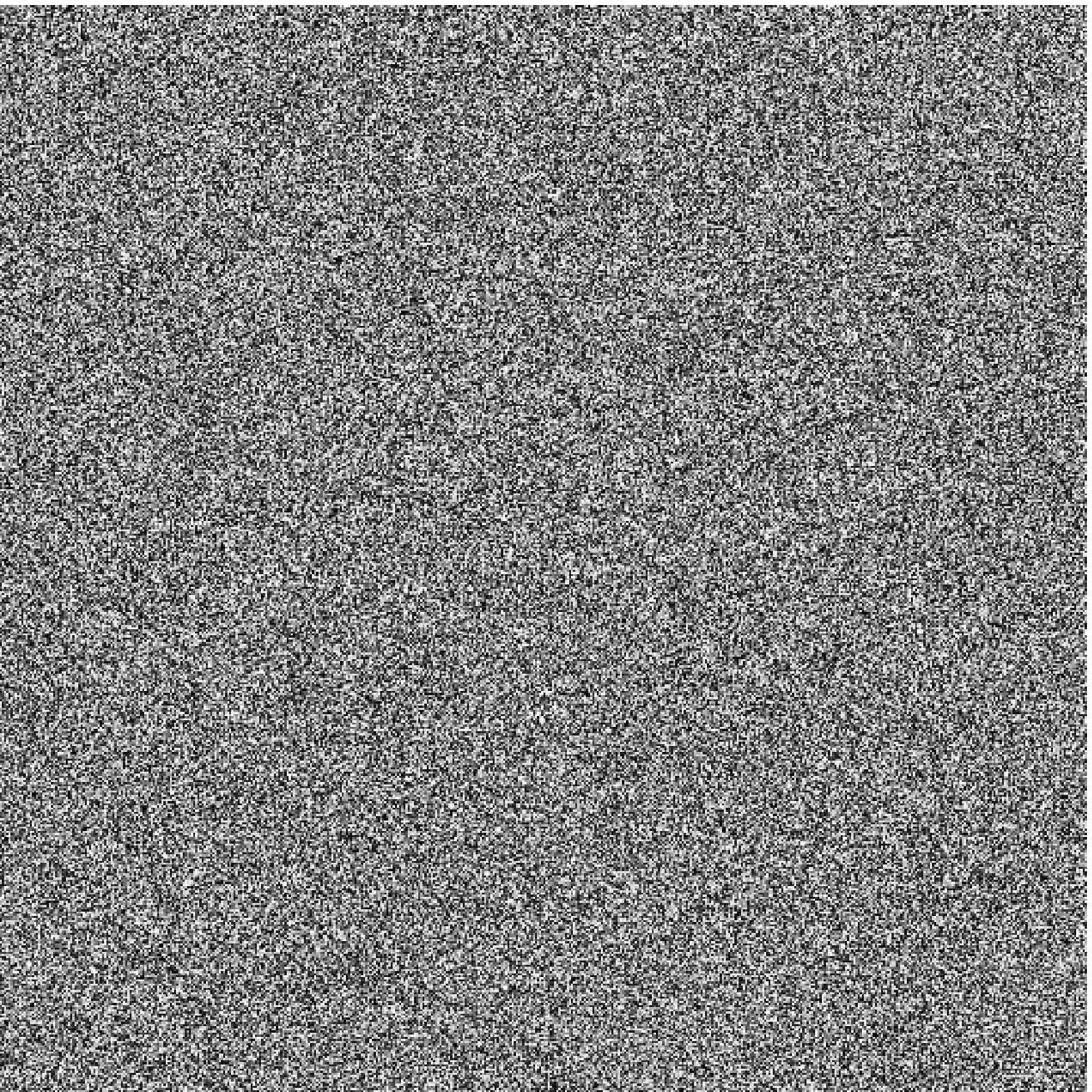}
\label{fig:dec_lena_wu}}
\caption{Decrypted images after compressing and decompressing
encrypted ones ($Q_{f}=90$)}
\label{fig:many_dec_img}
\end{figure}

Next, other conventional encryption methods are considered in terms of
compression performance. There are various encryption methods which
can maintain an image format after encrypting as well as the proposed
scheme. However, they are not suitable to EtC systems with JPEG
compression, because they do not consider using JPEG compression.
We numerically compared the encryption methods\cite{2016_Gaata,2012_wu} with the
proposed scheme. Figures \ref{fig:many_dec_img}\subref{fig:dec_lena_gaata} and \subref{fig:dec_lena_wu} indicate decrypted images after
compressing and decompressing encrypted ones, where Fig.
\ref{fig:many_dec_img}\subref{fig:orig_lena} is the original one. The image
quality of decrypted images heavily decreased due to JPEG compression as shown
in Figs. \ref{fig:many_dec_img}\subref{fig:dec_lena_gaata} and
\subref{fig:dec_lena_wu}, because they do not consider using JPEG
compression as well as most other conventional encryption methods. In contrast,
the image encrypted with the proposed scheme had almost same quality as that of
non-encrypted images. Note that Fig.
\ref{fig:many_dec_img}\subref{fig:dec_lena_wu} is the grayscale-based image\cite{2012_wu}, thus the PSNR value is not listed.

\begin{table*}
\centering
\caption{PSNR values of non-encrypted and decrypted images after uploading and downloading from Facebook. Boldface indicates highest score per $Q_{fu}$.}
\label{manipulation_facebook}
\begin{threeparttable}
\begin{tabular}{|c||c|c|c|c|c|c|c|}
\hline
\multirow{2}{*}{}                    &  \multicolumn{2}{c|}{Uploaded JPEG files} & \multicolumn{5}{c|}{$Q_{fu}$}                   \\ \cline{2-8} 
                                     &  Sub-sampling ratio &               Quantization table                   & 80    & 85    & 90    & 95    & 100   \\ \hline
\multirow{2}{*}{Non-encrypted}       & 4:2:0           &            \multirow{2}{*}{\shortstack{Luminance\\ Chrominance}}        & 31.004 & 30.228 & 32.105 & 32.364 & 32.409 \\ \cline{4-8} 
                                     & 4:4:4           &                   & 31.214 & 30.419 & 32.247 & 32.492 & 32.532 \\ \hline
\multirow{2}{*}{\shortstack{Conventional scheme\cite{kurihara2015encryption,KURIHARA2015}\\ ($B_{x}=B_{y}=16$)}} & 4:2:0               &        \multirow{2}{*}{\shortstack{Luminance\\ Chrominance}}        & 30.411\tnote{*} & 29.651\tnote{*} & 31.030\tnote{*} & 31.013\tnote{*} & 31.074\tnote{*} \\ \cline{4-8} 
                                     & 4:4:4 &                               & 31.068 & 30.220& 31.678& 31.650 & 31.712 \\ \hline
\multirow{2}{*}{\shortstack{Proposed scheme\\ ($B_{x}=B_{y}=8$)}}                &  \multirow{2}{*}{(Grayscale)}       &           Chrominance                & 31.786 & \textbf{32.289} & 32.629 & 33.422 & \textbf{33.807} \\ \cline{3-8} 
& &Luminance & \textbf{32.748} & 31.322 & \textbf{33.754} & \textbf{33.721} & \textbf{33.807} \\ \hline
\end{tabular}
\begin{tablenotes}
\item[*] Distorted by image manipulation on Facebook
\end{tablenotes}
\end{threeparttable}
\end{table*}

\begin{table*}
\centering
\caption{PSNR values of non-encrypted and decrypted images after uploading and downloading from Twitter. Boldface indicates highest score per $Q_{fu}$.}
\label{manipulation_twitter}
\begin{threeparttable}
\begin{tabular}{|c||c|c|c|c|c|}
\hline
\multirow{2}{*}{}                    & \multicolumn{2}{c|}{Uploaded JPEG files}& \multicolumn{3}{c|}{$Q_{fu}$}                   \\ \cline{2-6} 
                                     & Sub-sampling ratio  &              Quantization table                     & 90    & 95    & 100   \\ \hline
\multirow{2}{*}{Non-encrypted}       & 4:2:0           &            \multirow{2}{*}{\shortstack{Luminance\\ Chrominance}}        & 32.749 & 33.686 & 33.712 \\ \cline{4-6} 
                                     & 4:4:4           &                    & 33.100 & 33.627 & 33.659 \\ \hline
\multirow{2}{*}{\shortstack{Conventional scheme\cite{kurihara2015encryption,KURIHARA2015}\\ ($B_{x}=B_{y}=16$)}} & 4:2:0               &        \multirow{2}{*}{\shortstack{Luminance\\ Chrominance}}        & 32.612 & 33.578 & 33.605 \\ \cline{4-6} 
                                     & 4:4:4 &                             & 32.922 & 33.508 & 33.539 \\ \hline
\multirow{2}{*}{\shortstack{Proposed scheme\\ ($B_{x}=B_{y}=8$)}}                &  \multirow{2}{*}{(Grayscale)}       &           Chrominance               & \textbf{35.420} & 35.348 & \textbf{36.467} \\ \cline{3-6} 
& &Luminance  & 35.154 & \textbf{36.401} & \textbf{36.467} \\ \hline
\end{tabular}
\end{threeparttable}
\end{table*}

\subsection{Image Manipulation on Social Media}
We uploaded images encrypted by using the proposed scheme to Facebook and
Twitter, and then downloaded them, as well as images encrypted using the
conventional scheme\cite{kurihara2015encryption,KURIHARA2015} and non-encrypted
images to confirm the effectiveness of the proposed scheme.
\subsubsection{Experimental Conditions}
In our experiments, the same dataset and parameters for JPEG compression as in Sec.\,\ref{single_compress} were utilized for the evaluation.
The following procedure was carried out according to Fig.\,\ref{fig:etc}.
\begin{itemize}
\item[1)]Generate encrypted image $I_{e}$ from original image $I$.
\item[2)]Compress encrypted image $I_{e}$ with $Q_{fu}$.
\item[3)]Upload encrypted JPEG image $I_{ec}$ to SNS providers.
\item[4)]Download recompressed JPEG image $\hat{I_{ec}}$ from the providers.
\item[5)]Decompress encrypted JPEG image $\hat{I_{ec}}$.
\item[6)]Decrypt manipulated image $\hat{I_{e}}$.
\item[7)]Compute the PSNR value between original image $I$ and decrypted image $\hat{I}$.
\end{itemize}
To compare PSNR values in step 7, original image $I$ was compressed without
any encryption and then uploaded and downloaded.
The downloaded images were decompressed, and
then, the average PSNR values for 1338 images per $Q_{fu}$ were calculated.
\par
We focused on Facebook and Twitter, which always recompress uploaded images when
the images meet the conditions for $Q_{fu}$ as shown in
Table\,\ref{tb:change_sampling}.
\subsubsection{Experimental Results}
Table\,\ref{manipulation_facebook} shows the experimental result of Facebook.
As described in Sec.\,\ref{subsampling}, the images decrypted with the
conventional scheme and 4:2:0 sub-sampling ratio included block distortion
like in Fig\,.\ref{fig:oriencex}(a).
Therefore, the PSNR values of the decrypted images were much lower than the
others.
In comparison, images decrypted by using the proposed scheme always avoided
the interpolation on Facebook, so the PSNR values of decrypted images were
higher than the others.
\par
Table\,\ref{manipulation_twitter} shows the result for Twitter.
As indicated in the Table, a similar tendency to
Facebook was shown.
While Facebook recompresses all uploaded images by users, Twitter recompresses uploaded images under some conditions such as $Q_{fu}\geqq85$ for 4:2:0 sub-sampling ratio.
For this reason, we uploaded images compressed with $Q_{fu}=90,95,100$.
Twitter recompresses uploaded images with 4:2:0 sub-sampling in the DCT
domain, thereby generating no distortion\cite{CHUMAN2017APSIPA}.
However, interpolation is carried out by users when users download and
decompress the images.
Therefore, the PSNR values of the decrypted images with 4:2:0 sub-sampling
ratio were lower than with the proposed scheme.
\par
As shown in Tables\,\ref{manipulation_facebook} and \ref{manipulation_twitter},
most PSNR values of the images decrypted with the proposed scheme using
the luminance table were higher than with the chrominance one.
It has been known that most SNS providers use the luminance table for the uploaded grayscale images in recompression.
That is to say, uploaded images compressed with the luminance table by users are recompressed with the luminance table again by SNS providers.
As a result, using the luminance table would decreases the quantization
error in recompression.
\subsection{Robustness against Jigsaw Puzzle Solver Attacks}
Next, the security of grayscale-based block scrambling image encryption against the jigsaw puzzle solver was evaluated. 
\subsubsection{Experimental Conditions}
We assembled encrypted images by using the extended jigsaw puzzle solver\cite{CHUMAN2017ICASSP,CHUMAN2017ICME}.
The following three measures\cite{Gallagher_2012_CVPR}\cite{Cho_2010_CVPR} were used to evaluate the results.
\\
{\bf Direct comparison ($D_{c}$):} represents the ratio of the number of pieces which are in the correct position. $Dc$ for image $I_d$, namely, $Dc(I_d)$ is calculated as
\begin{IEEEeqnarray}{ll}
	D_{c}(I_d)\ &=\frac{1}{n} \sum_{i=1}^{n}d_{c}(i), \\
	d_{c}(i)&=
	\left\{
	\begin{array}{ll}
		1, & {\rm if \;} I_{d}(i) {\rm \; is \; in \; the \; correct \; position} \nonumber \\
		0, & {\rm otherwise}
	\end{array}
	\right.
\end{IEEEeqnarray}
where $I_d(i)$ represents the position of a piece $i$ in image $I_d$
\\
{\bf Neighbor comparison ($N_{c}$):} is the ratio of the number of correctly joined blocks. $Nc$ for image $I_d$, namely, $Nc(I_d)$ is calculated as
\begin{IEEEeqnarray}{lll}
	Nc(I_d)&=\frac{1}{B}\sum_{k=1}^{B}n_c(k), \\
	n_{c}(k)&=
	\left\{
	\begin{array}{ll}
		1, & {\rm if \;} b_{k}{\rm \; is \; joined \; correctly} \nonumber \\
		0, & {\rm \; otherwise}
	\end{array}
	\right.
\end{IEEEeqnarray}
where $B$ is the number of boundaries among pieces in $I_d$, and $b_k$ is the $k$th boundary in $I_d$. 
For an image with $u \times v$ blocks, there are $B=2uv-u-v$ boundaries in the image.
\\
{\bf Largest component ($L_{c}$):} is the ratio of the number of the
largest joined blocks that have correct adjacencies to the
number of blocks in an image. $Lc$ for image $I_d$, namely, $Lc(I_d)$ is calculated as
\begin{IEEEeqnarray}{ccc}
	L_{c}(I_d) = \frac{1}{n}\max_{j} \{ l_c(I_{d},j)\}, j=1,2,\cdots,m
\end{IEEEeqnarray}
where $l_{c}(I_d,j)$ is the number of blocks in the $j$th partial correctly assembled area,
 and $m$ is the number of partial correctly assembled areas.
\par
In the measures, $D_{c}, N_{c}, L_{c} \in [0,1]$, a larger value means a higher compatibility.
\par
We used 20 images randomly chosen from the UCID dataset, and each image was
resized to 256$\times$192.
Forty different encrypted images were generated from one ordinary image by
using different keys.
We assembled the encrypted images by using the jigsaw puzzle solver and chose the image that had the highest sum of $D_{c}$, $N_{c}$, and $L_{c}$.
We performed this procedure for each encrypted image independently and
calculated the average $D_{c}$, $N_{c}$, and $L_{c}$ for the 20 images.
\begin{figure*}[t]
\captionsetup[subfigure]{justification=centering}
\centering
\begin{center}
\subfloat[Original image\newline ($X\times Y =256\times 192$)]
{\includegraphics[height=3.5cm]{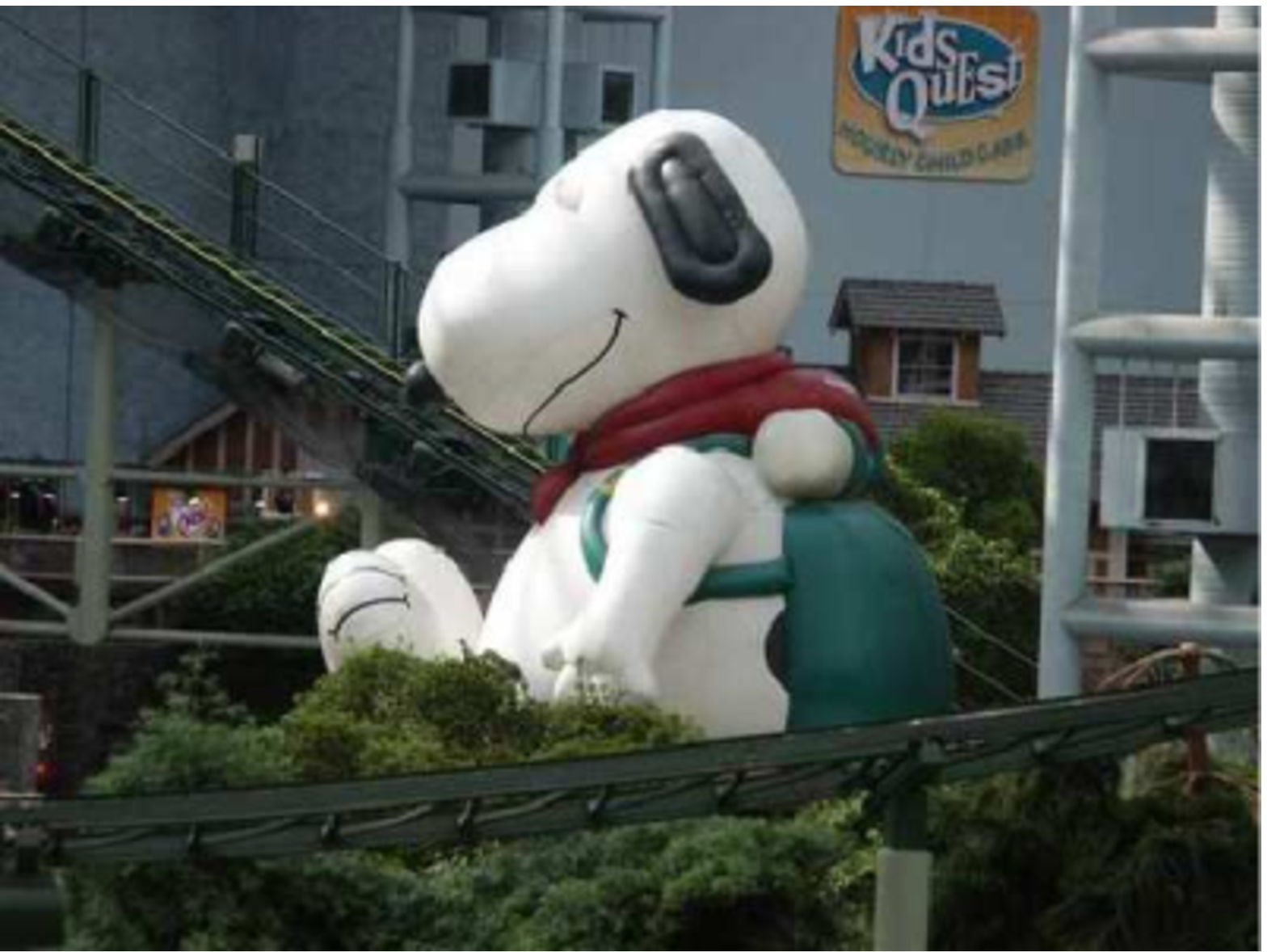}
\label{fig:ass_ori}}
\subfloat[Assembled image\newline (Conventional scheme, $B_{x}=B_{y}$= 16, $D_{c}=0, N_{c}=0.111, L_{c}=0.078$)]
{\includegraphics[height=3.5cm]{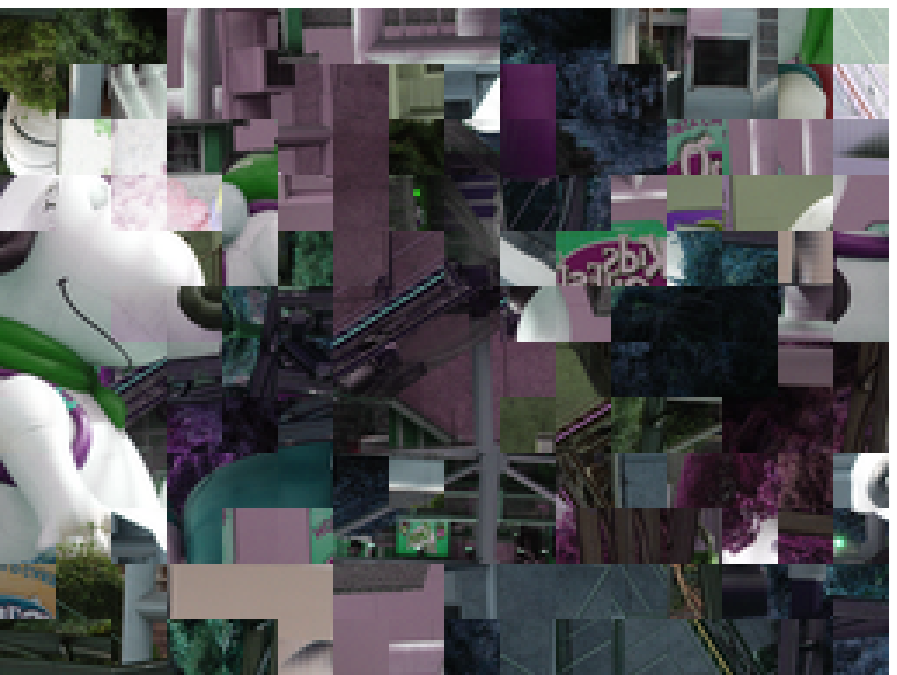}
\label{fig:ass_conv16}}
\subfloat[Assembled image\newline (Grayscale, $B_{x}=B_{y}$ = 16,\newline $D_{c}=0, N_{c}=0.042, L_{c}=0.036$)]
{\includegraphics[height=3.5cm]{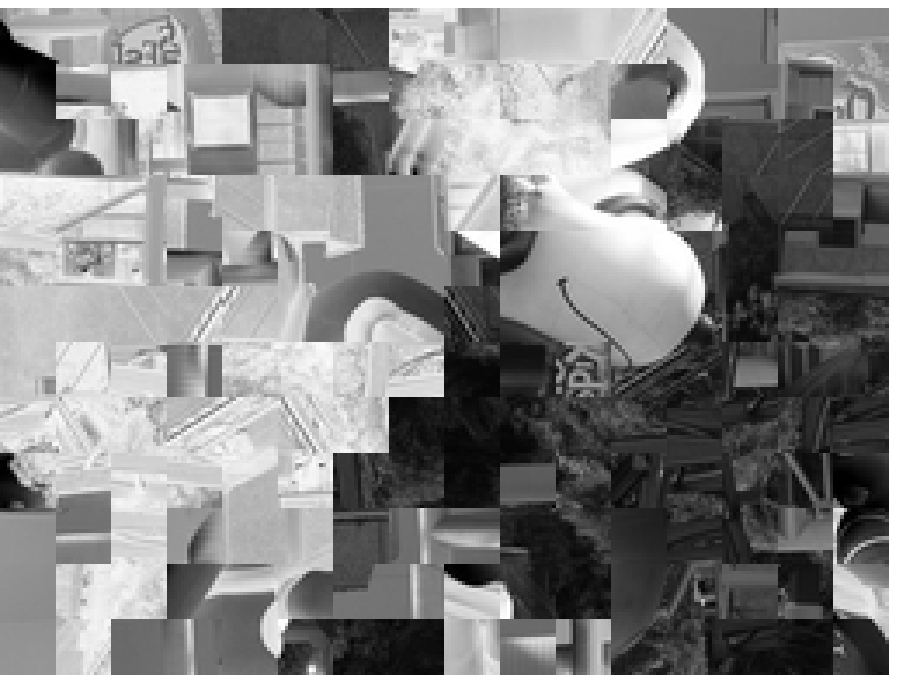}
\label{fig:ass_gray16}}
\end{center}
\caption{Images assembled from encrypted ones with the conventional scheme. The image (c) is assembled from grayscale-based encrypted image with one channel.}
\label{fig:assembled_conv}
\end{figure*}
\begin{figure*}[t]
\captionsetup[subfigure]{justification=centering}
\centering
\begin{center}
\subfloat[Assembled image\newline (Proposed, $B_{x}=B_{y}$ = 32,\newline $D_{c}=0.001, N_{c}=0.000, L_{c}=0.021$)]
{\includegraphics[height=3.5cm]{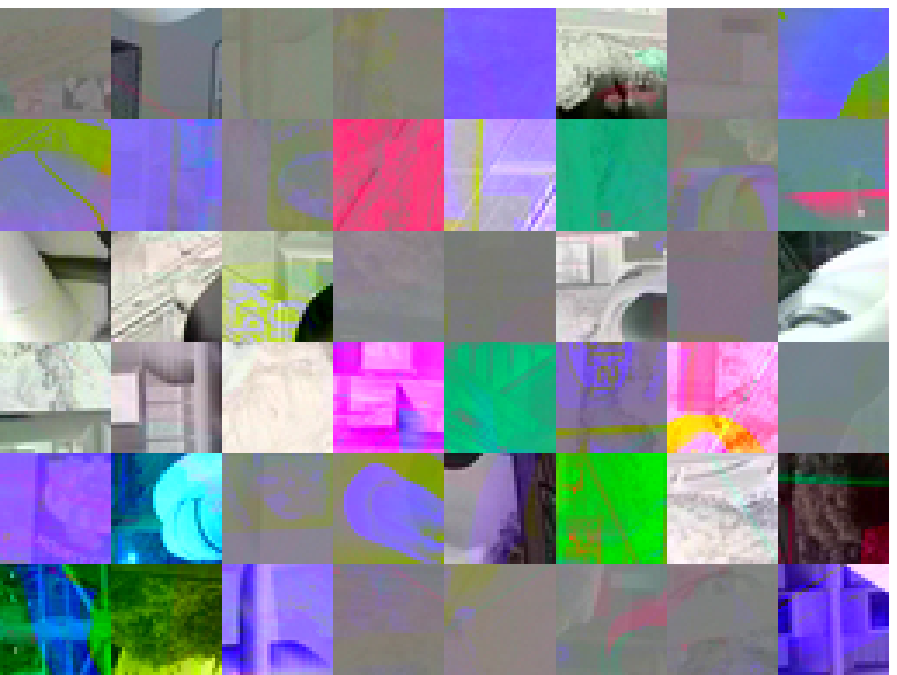}
\label{fig:ass_pro32_1}}
\subfloat[Assembled image\newline (Proposed, $B_{x}=B_{y}$ = 16,\newline $D_{c}=0.001, N_{c}=0.001, L_{c}=0.006$)]
{\includegraphics[height=3.5cm]{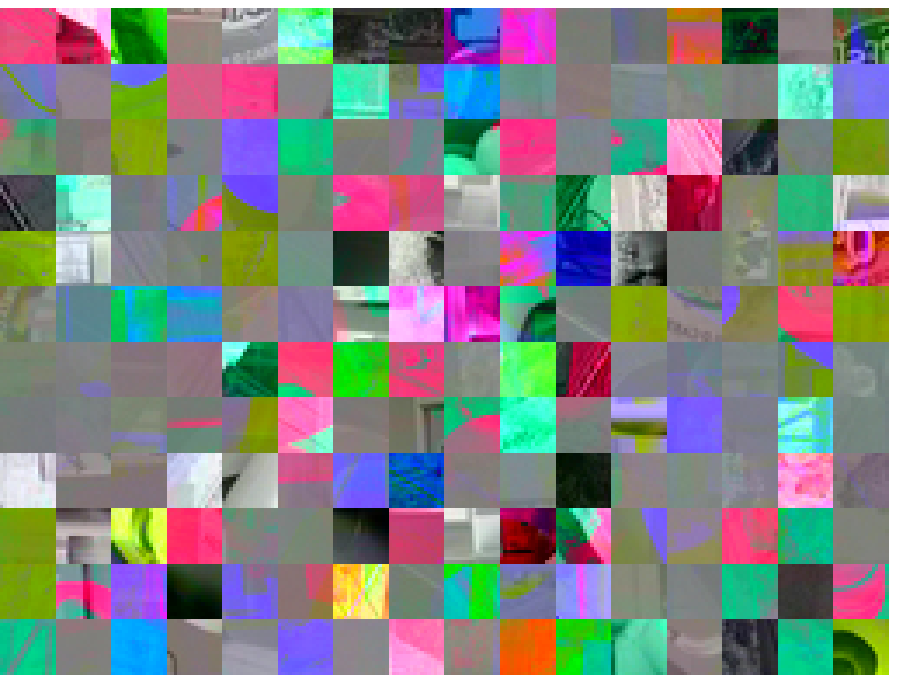}
\label{fig:ass_pro16_1}}
\subfloat[Assembled image\newline (Proposed, $B_{x}=B_{y}$ = 8,\newline $D_{c}=0.001, N_{c}=0.001, L_{c}=0.002$)]
{\includegraphics[height=3.5cm]{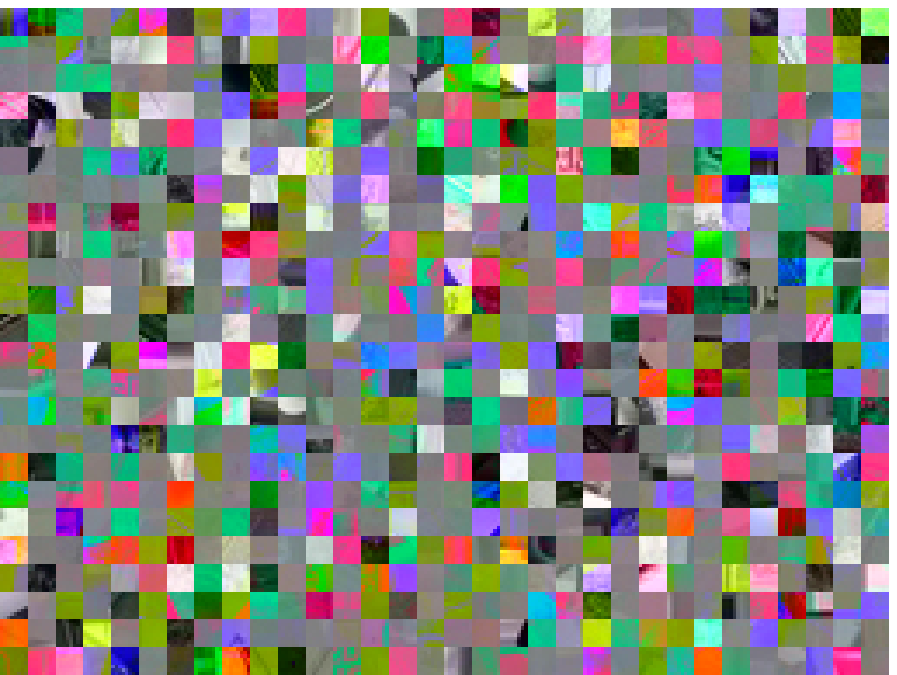}
\label{fig:ass_pro8_1}}
\\
\end{center}
\caption{Images assembled from encrypted ones with the proposed scheme}
\label{fig:assembled_pro}
\end{figure*}
\subsubsection{Experimental results}
Table\,\ref{jigsaw_result} shows robustness against an extended jigsaw puzzle solver attack\cite{CHUMAN2017ICASSP,CHUMAN2017ICME}.
To evaluate the effect of reducing the number of color channels on puzzle assembly, images with only one channel, namely grayscale-based images, were encrypted in accordance with Step 5 in Sec.\,\ref{new} and then assembled.
Examples of images assembled from encrypted ones with the proposed and conventional scheme are shown in Figs.\,\ref{fig:assembled_conv}\subref{fig:ass_conv16} and \subref{fig:ass_gray16}; Fig.\,\ref{fig:assembled_conv}\subref{fig:ass_ori} shows the original one.
As shown in Fig.\,\ref{fig:assembled_conv}\subref{fig:ass_conv16}, it is possible to partially assemble encrypted images with the conventional scheme if the number of blocks is small.
Since the scores for the encrypted images with one color channel were lower than
those of the ones assembled with the conventional scheme, reducing the number of
color channels makes assembling puzzles more difficult.
\par
Figures\,\ref{fig:assembled_pro}\subref{fig:ass_pro32_1},
\subref{fig:ass_pro16_1} and \subref{fig:ass_pro8_1} show the examples of
images assembled from encrypted ones with the proposed scheme, where Fig.\,\ref{fig:assembled_conv}\subref{fig:ass_ori} is the original image.
As shown in Table\,\ref{jigsaw_result} and Figs.\,\ref{fig:assembled_pro}\subref{fig:ass_pro32_1}, \subref{fig:ass_pro16_1} and \subref{fig:ass_pro8_1}, the scores for the proposed scheme were much lower than the other scores.
Even when large block sizes ($B_{x}=B_{y}=16$) were used for the proposed
encryption, the scores of assembled images were far lower than with
the conventional scheme with $L_{c}=0.021$.
This is because images encrypted with the proposed scheme have a large number of encrypted blocks and less color information in the blocks.
The proposed scheme thus has higher security against jigsaw puzzle solver attacks than the conventional scheme.

\begin{table}[t]
\centering
\caption{Security evaluation of the conventional and proposed scheme against the extended jigsaw puzzle solver}
\label{jigsaw_result}
\begin{tabular}{c|c|c|c|c|c|}
\cline{1-6}
                                                   
\multicolumn{1}{|c||}{\multirow{2}{*}{Encryption type}}             & \multirow{2}{*}{\shortstack{Color\\ channel}} & \multirow{2}{*}{Block size} & \multirow{2}{*}{$D_{c}$} & \multirow{2}{*}{$N_{c}$} & \multirow{2}{*}{$L_{c}$} \\
\multicolumn{1}{|c||}{}                              &                             &                                &                     &                     &                     \\ \hline
\multicolumn{1}{|c||}{\multirow{2}{*}{\shortstack{Conventional\\ scheme\cite{kurihara2015encryption,KURIHARA2015}}}} & \multirow{2}{*}{RGB}                       & $32\times32$                            & 0.592               & 0.464               & 0.604               \\ \cline{3-6} 
\multicolumn{1}{|c||}{}                              &                        & $16\times16$                            & 0.005               & 0.111               & 0.120               \\ \hline
\multicolumn{1}{|c||}{\multirow{2}{*}{Grayscale}}    & \multirow{2}{*}{Grayscale}                      & $32\times32$                             & 0.546              & 0.532               & 0.597               \\ \cline{3-6} 
\multicolumn{1}{|c||}{}                              &                        & $16\times16$                            & 0.002               & 0.062               & 0.066               \\ \hline
\multicolumn{1}{|c||}{\multirow{3}{*}{\shortstack{Proposed\\ scheme}}}     & \multirow{3}{*}{Grayscale}                      & $32\times32$                            & 0.001               & 0.000               & 0.021               \\ \cline{3-6} 
\multicolumn{1}{|c||}{}                              &                       & $16\times16$                             & 0.001               & 0.001               & 0.006               \\ \cline{3-6} 
\multicolumn{1}{|c||}{}                              &                       & $8\times8$                            & 0.001               & 0.001               & 0.002               \\ \hline
\end{tabular}
\end{table}
\begin{figure}[t]
\centering
\includegraphics[width =8cm]{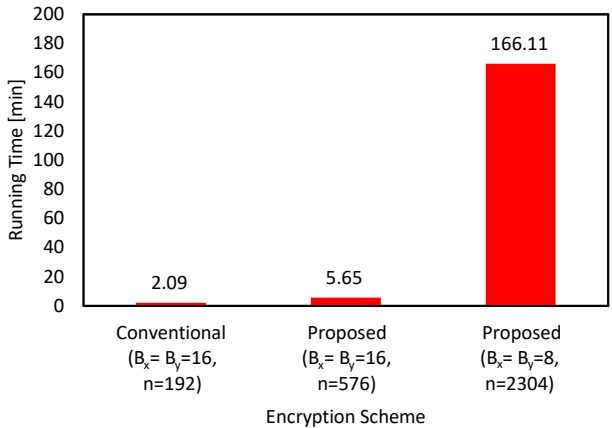}
\caption{Running time of assembling encrypted images from resized UCID dataset
($256 \times 192$)}
\label{fig:com_time}
\end{figure}
\subsubsection{Running time to assemble jigsaw puzzles}
Figure\,{\ref{fig:com_time}} shows the running time to assemble encrypted
images by using the jigsaw puzzle solver\cite{CHUMAN2017ICASSP,CHUMAN2017ICME},
where the average time of 15 images from resized UCID dataset were plotted. We
compared the running time to assemble images encrypted with the conventional scheme ($B_x=B_y=16$) and the proposed one ($B_x=B_y=16$ and $B_x=B_y=8$). The jigsaw puzzle solver
was implemented in MATLAB2015a on a PC with a 3.3GHz processor and a main
memory 16Gbytes (Processor:Intel Core i7-5820K 3.3GHz, OS:Ubuntu 16.04 LTS).
\par As shown in Fig.\,\ref{fig:com_time}, although the images encrypted
by using the proposed method ($B_x=B_y=8$) were solved in 166.11 minutes, the
scores of assembled images were very low as $L_c=0.002$ (See Table.\,\ref{jigsaw_result}). It obviously takes
more time to assemble encrypted images than that of images encrypted
by the conventional scheme. The reason is that the proposed scheme can
offer a smaller block size, the larger number of blocks and less color
information.  As a result, the proposed scheme can enhance security against
ciphertext-only attacks in terms of both computational complexity and the
accuracy of assembled results.

\section{Conclusion}
\label{sec:conclusion}
We proposed a novel block-scrambling image encryption scheme that enhances
the security of EtC systems for JPEG images.
Although $B_{x}=B_{y}=16$ is used as the smallest block size in the conventional scheme to avoid the effect of color sub-sampling, the proposed scheme enables us to use $B_{x}=B_{y}=8$ as a block size, which enhances robustness against ciphertext-only attacks.
Although jigsaw puzzle solver attacks can be assumed to occur as cipher-text
only attacks, images encrypted with the proposed scheme include a larger
number of small blocks, which makes assembling encrypted images much more
difficult.
In comparison, decrypted images with the conventional scheme sometimes
include some block distortion due to the interpolation on social media.
The proposed scheme makes it possible to avoid the effect of the
interpolation on social media due to the use of grayscale-based images. As a result, the proposed scheme has a better performance than the conventional one in terms of the image quality.
Experimental results showed the EtC systems with the proposed scheme are applicable to Twitter and Facebook.
The proposed scheme is also applicable to other SNS providers and cloud
photo services like Tumblr, iCloud and Google Photos.
In addition, the robustness of the proposed scheme against jigsaw puzzle solver
attacks was confirmed in the experiment.
\section*{Acknowledgment}
This work was partially supported by Grant-in-Aid for Scientific Research(B), No.17H03267, from the Japan Society for the Promotion Science.
\ifCLASSOPTIONcaptionsoff
  \newpage
\fi

\bibliographystyle{IEEEtran}

\begin{IEEEbiography}[{\includegraphics[width=1in,height=1.25in,clip,keepaspectratio]{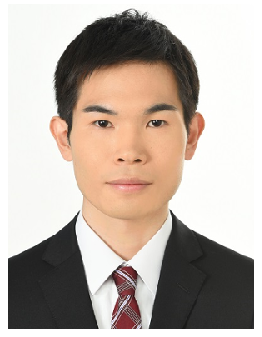}}]{Tatsuya Chuman}
received his B.Eng. degree from Toyo University, Japan in 2016.
From 2016, he has been a Master course student at Tokyo Metropolitan University.
His research interests are in the area of image processing.
\end{IEEEbiography}

\begin{IEEEbiography}[{\includegraphics[width=1in,height=1.25in,clip,keepaspectratio]{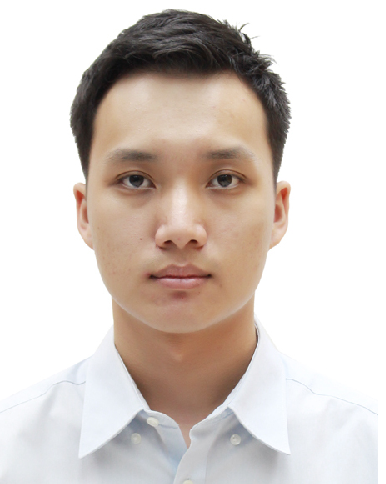}}]{Warit
Sirichotedumrong}received his B.Eng. and M.Eng. degrees from King Mongkut's University of Technology Thonburi, Thailand in 2014 and 2017, respectively.
From 2017, he has been a Doctor course student at Tokyo Metropolitan University.
His research interests are in the area of image processing.
\end{IEEEbiography}
\begin{IEEEbiography}[{\includegraphics[width=1in,height=1.25in,clip,keepaspectratio]{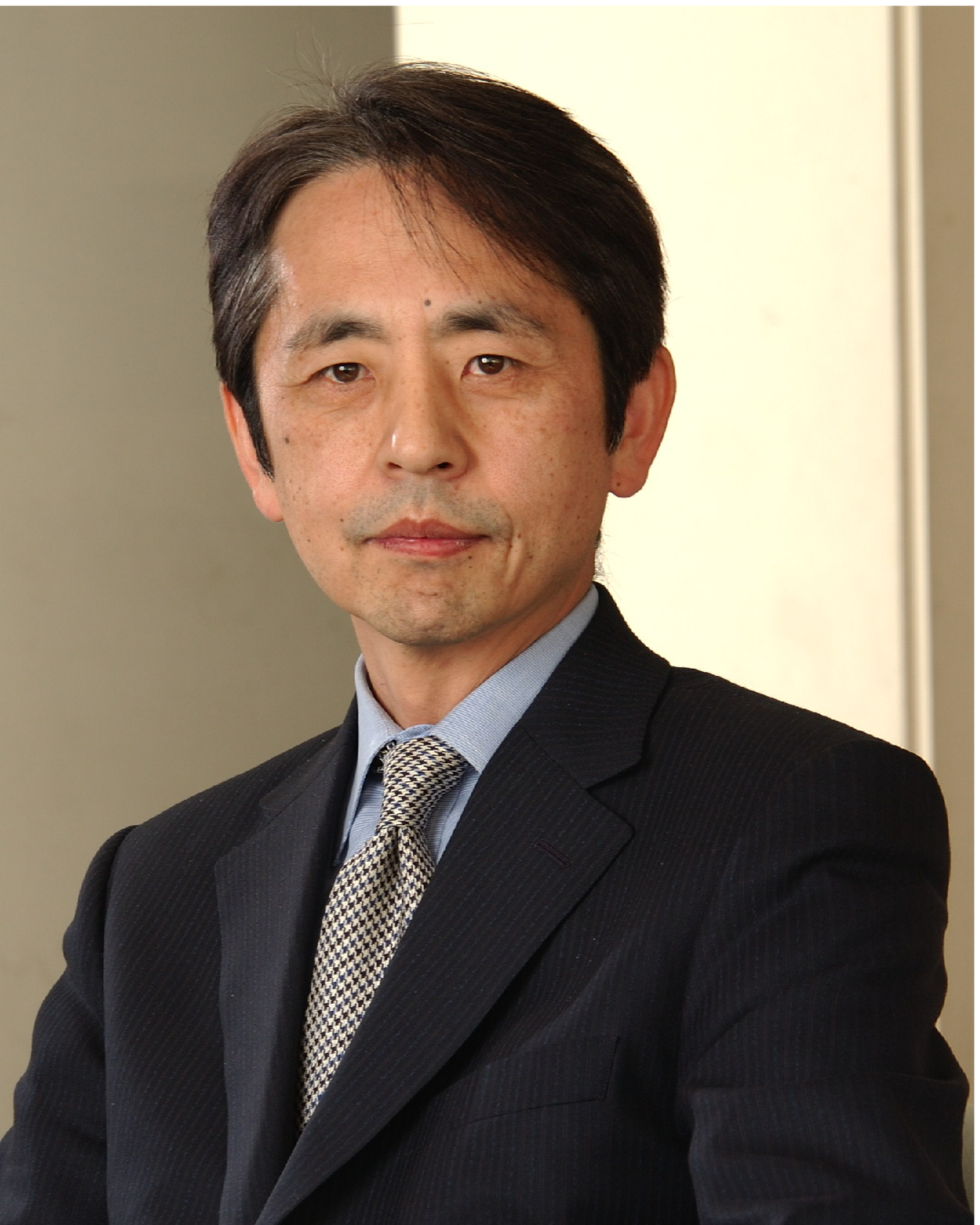}}]{Hitoshi
Kiya}received his B.E and M.E. degrees from Nagaoka University of Technology, in 1980 and 1982 respectively, and his Dr. Eng. degree
from Tokyo Metropolitan University in 1987. In 1982, he joined Tokyo
Metropolitan University, where he became a Full Professor in 2000.
From 1995 to 1996, he attended the University of Sydney, Australia as
a Visiting Fellow. He is a Fellow of IEEE, IEICE and ITE. He currently
serves as President-Elect of APSIPA, and he served as Regional
Directorat-Large for Region 10 of the IEEE Signal Processing Society
from 2016 to 2017. He was also President of the IEICE Engineering
Sciences Society from 2011 to 2012, and he served there as a Vice
President and Editor-in-Chief for IEICE Society Magazine and Society
Publications. He was Editorial Board Member of eight journals,
including IEEE Trans. on Signal Processing, Image Processing, and
Information Forensics and Security, Chair of two technical committees
and Member of nine technical committees including APSIPA Image, Video,
and Multimedia Technical Committee (TC), and IEEE Information
Forensics and Security TC. He has organized a lot of international
conferences, in such roles as TPC Chair of IEEE ICASSP 2012 and as
General Co-Chair of IEEE ISCAS 2019. He has received numerous awards,
including six best paper awards.
\end{IEEEbiography}

\end{document}